\documentclass[aps,pre,superscriptaddress,longbibliography]{revtex4-2}
\usepackage{array}
\usepackage{amsmath,amssymb,amsfonts}
\usepackage{tabularx}
\usepackage{graphicx}
\usepackage{dcolumn}       
\usepackage{bm}            
\usepackage{hyperref}
\usepackage{xcolor}
\hypersetup{hidelinks}
\usepackage{booktabs}
\usepackage{multirow}
\usepackage{siunitx}       
\usepackage{microtype}     
\usepackage{float}         
\usepackage[section]{placeins} 
\usepackage{flafter}       

\newcommand{\Hadm}{H_\mathrm{Admin}}
\newcommand{\Hgeo}{H_\mathrm{Geo}}
\newcommand{\Htrd}{H_\mathrm{Trade}}
\newcommand{\Hfull}{H_\mathrm{Full}}
\newcommand{\Hcomb}{H_\mathrm{comb}}
\newcommand{\Hstar}{H^*}
\newcommand{\wlcp}{w_\mathrm{LCP}}
\newcommand{\Vadm}{V_\mathrm{Admin}}
\newcommand{\Vgeo}{V_\mathrm{Geo}}
\newcommand{\Vtrd}{V_\mathrm{Trade}}
\newcommand{\ns}{\text{n.s.}}

\newcommand{\tstar}{t^*}

\newcommand{\DHH}{\Delta H}
\newcommand{\gap}{\tau}

\definecolor{adminred}{RGB}{216,90,48}
\definecolor{geoblue}{RGB}{55,138,221}
\definecolor{tradegr}{RGB}{29,158,117}
\definecolor{hstargold}{RGB}{186,117,23}

\begin{document}

\title{Topological Signatures of Imperial Collapse and Fragmentation:\\
Administrative Dissolution, Territorial Reorganization,\\
and Early-Warning Observables in the Han Dynasty Network\\
(206~BCE\,--\,220~CE)}

\author{Jos\'e de Jes\'us Bernal-Alvarado}
\email{bernal@ugto.mx}
\affiliation{Physics Engineering Department, Universidad de Guanajuato, M\'{e}xico}

\author{David Delepine}
\email{delepine@ugto.mx}
\affiliation{Physics Department, Universidad de Guanajuato, M\'{e}xico}

\author{Carlos Pinedo Guadarrama}
\email{c.pinedoguadarrama@ugto.mx}
\affiliation{Physics Department, Universidad de Guanajuato, M\'{e}xico}

\date{\today}

\begin{abstract}
We extend the persistent homology formalism done in references~\cite{paper1,paper2} to the Han Dynasty ($206~\mathrm{BCE}$--$220~\mathrm{CE}$),
testing whether the collapse threshold $\Hstar = 0.5241$ and
the three-network decomposition methodology established for the
Roman--Byzantine case generalize to a mechanistically distinct
imperial system.
Three complementary networks are constructed from geospatial and
historical data: an administrative network ($\Hadm$, 392~prefectures
from CHGIS~v6), a geographic network ($\Hgeo$, with node end-dates
extended beyond administrative dissolution), and an 81-node Silk
Road trade network ($\Htrd$) as a structural control.
Edge costs are derived via least-cost path analysis on a
$1~\mathrm{km/pixel}$ digital elevation model, modulated by five
historical friction channels.
$\Hadm$ collapses to zero
at 220~CE (phase~G slope $-0.0261~\mathrm{yr}^{-1}$,
$R^2 = 0.689$) while $\Hgeo$ simultaneously \emph{increases}
to $3.065$, producing a divergence $\Delta H = 3.065$---the
topological signature of \emph{internal fragmentation}.
The cross-network Wasserstein distance $W_\times(\mathrm{Admin},\mathrm{Geo})$
jumps from exactly zero to $737$ at $d=190$~CE, one decade after
the Yellow Turban Rebellion, and three long-lived $\beta_1$ cycles
matching the Wei, Shu, and Wu domains emerge under the post-184~CE
fragmentation model---thirty years before formal partition.
Three independent early-warning indicators (Wasserstein velocity
$\dot{W}_2$, correlation length $\xi$, and the Integrated Change
Tracker) signal collapse onset 45--50~years before formal
dissolution.

\end{abstract}


\maketitle

\section{\label{sec:intro}Introduction}

The collapse of large-scale political systems is among the most
consequential and least understood phenomena in  History.
From Tainter's metabolic model of diminishing returns on
complexity~\cite{Tainter1988} to Turchin's structural-demographic
cycles~\cite{Turchin2003}, theoretical models for imperial collapse
remain largely qualitative or dependent on system-specific proxies.
A quantitative, mechanism-agnostic measure of systemic
disintegration would provide an objective basis for
cross-civilizational comparison.
Topological data analysis (TDA), and persistent homology in
particular, provides a candidate measure~\cite{Edelsbrunner2002,
Zomorodian2005}: given the matrix of travel costs between
all pairs of settlements, it extracts a single number---the
persistent entropy $H$~\cite{Rucco2017,Chintakunta2015}---that
quantifies how many independent, redundant routes the network
contains. 
In this paper, we shall extend the TDA analysis to the period of China history  ($206~\mathrm{BCE}$--$220~\mathrm{CE}$) corresponding to the Han Dynastie period.

The Han collapse was essentially internal:
The Yellow Turban Rebellion of 184~CE ignited a cascade of warlord
fragmentation that dissolved the unified state from within.  
At the end of its period,  the Han empire had split into the Three Kingdoms (Wei, Shu,
and Wu)---a partition of \emph{authority}, not of territory.


Applying the TDA tools to Han Dynasty, we can summarize our  main results as followed:

\begin{enumerate}

  \item \textbf{Three-network decomposition.}
  We construct three independent networks---$\Hadm$, $\Hgeo$,
  and $\Htrd$---from primary geospatial data.

  \item \textbf{Topological signature of internal fragmentation.}
  $\Hgeo$ \emph{increases} while $\Hadm$ collapses,
  with the divergence $\Delta H(t) = \Hgeo(t) - \Hadm(t)$
  rising from zero at 180~CE to $+3.065$ at the end of the period.
  The cross-network Wasserstein distance
  $W_\times(\mathrm{Admin}, \mathrm{Geo})$ confirms this in
  the full barcode metric.
  The Three~Kingdoms emerge as $\beta_1$ cycles with a
  $10.9\times$ signal-to-noise gap at $d = 190$~CE---thirty years
  before formal partition under the post-184~CE fragmentation model.

  \item \textbf{Early-warning observables and collapse chronology.}
  Three independent indicators---Wasserstein velocity $\dot{W}_2$,
  correlation length $\xi$, and the Integrated Change Tracker
  (ICT)---provide an early-warning chronology.
  ICT first exceeds its threshold at $d = 170$~CE, 50~years before
  dissolution and at a moment when $\Hadm$ alone shows no
  significant trend.
  $\dot{W}_2$ exceeds $4\times$ baseline at $d = 175$~CE,
  45~years before dissolution.
  The correlation length $\xi_\mathrm{Geo}$ diverges monotonically
  to $55\times$ baseline by 220~CE.

  \item \textbf{Commercial-network stability.}
  $\Htrd$ is  stable across all 420~years
  suggesting that long-distance trade circuits constitute a
  topological pattern invariant to the political systems
  that facilitate them.


\end{enumerate}

The remainder of the paper is organized as follows.
Section~\ref{sec:data} describes data sources and network construction.
Section~\ref{sec:friction} summarizes the five-channel friction model
(full specifications are given in Sec.~\ref{sec:friction}).
Section~\ref{sec:tda} describes the TDA pipeline
(full specification in Ref.~\cite{paper2}).
Section~\ref{sec:networks} defines the three networks and
$\Hcomb$.
Section~\ref{sec:results} presents the main results, including
the Admin--Geo divergence ,
the Three~Kingdoms $\beta_1$ identification, and the
early-warning observables.
Section~\ref{sec:discussion} synthesizes findings. 




\section{\label{sec:data}Data and network construction}
 
As a results of a geopspatial analysis process explained below,   a weighted graph $G_d = (V_d, E_d, w_d)$ is produced for each
of the 43~decades $d \in \{-200, -190, \ldots, 220\}$~CE where $V_d$ corresponds to the network nodes, $E_d$ to the edges associates to the node. 
Edge weights $w_d$ encode the traversal cost of inter-node routes
as inferred from real topography, modulated by five
historical friction channels described in Sec.~\ref{sec:friction}.
\subsection{\label{sec:chgis}Administrative nodes:
            China Historical GIS v6}
 
Administrative nodes are drawn from the
\emph{China Historical GIS} version~6
(CHGIS~v6)~\cite{CHGIS}, a georeferenced dataset of
2,755 administrative entities active during the studied period,
each with precise geodetic coordinates (WGS84),
and exact start (\texttt{BEG\_YR}) and end (\texttt{END\_YR})
years derived from the primary chronicles (\textit{Han Shu},
\textit{Hou Han Shu})~\cite{HanShu,HouHanShu}.
 Three hierarchical levels
relevant to our analysis can be observed:
 
\begin{itemize}
  \item \textit{Guo}~(commandery prefecture):
        170~nodes, administrative level~5.
        The primary unit of Han territorial governance;
        each was headed by an imperially appointed prefect
        (\textit{taishou}).
  \item \textit{Houguo}~(marquis state):
        222~nodes, administrative level~5.
        Semi-autonomous sub-entities within the Guo system,
        with hereditary lords (\textit{hou}) designated by
        the Han throne.
  \item \textit{Xian}~(county):
        2,337~nodes, administrative level~6.
        Sub-prefectural units. 
\end{itemize}
 
Together, the 392~Guo and Houguo nodes constitute the node set
for both the administrative network ($\Hadm$, Sec.~\ref{sec:Hadm})
and the geographic network ($\Hgeo$, Sec.~\ref{sec:Hgeo}).
The  CHGIS dataset informs us on  the
\texttt{BEG\_YR}/\texttt{END\_YR} used to include or not the node in the corresponding network. 
When CHGIS marks an entity as ended, it records that the
administrative unit was abolished, merged, or lost to
imperial control.
 
The number of active administrative nodes varies from
$n_\mathrm{min}=19$ (220~CE, terminal Han) to
$n_\mathrm{max}=145$ ($-10$~CE, late Western Han peak),
closely tracking the documented expansion and contraction
of the Han prefectural system.
 
\subsection{\label{sec:silkroad}Exterior trade nodes:
            Silk Road dataset}
 
Exterior trade nodes are taken from a georeferenced
Silk Road dataset~\cite{SilkRoad} comprising 83~waypoints
and 88~segments, with documented activity dates covering the
full Han period.
After intersection with the DEM domain (see Sec.~\ref{sec:dem}),
81~nodes and 86~edges are retained during all the period.
Oasis cities such as Dunhuang, Khotan,
and Kashgar persisted across political transitions.
 
Edge costs for the Silk Road segments are computed by the
same least-cost path (LCP) algorithm used for interior edges
(Sec.~\ref{sec:lcp}), operating on the same cost surface.
 
\subsection{\label{sec:dem}Digital elevation model}
 
The terrain model is a $6299 \times 6170$~pixel raster
at $\Delta x = \Delta y = 1~\mathrm{km/pixel}$ resolution,
covering continental China and the Tarim Basin.
Nodata pixels (value $-32768$) and below-sea-level
pixels are flagged as impassable.
The Digital Elevation Model (DEM) is the critical component that allows edge costs
to be derived from real topography rather than
historical estimates~\cite{Verhagen2013}:
mountain passes, river valleys, and high-altitude plateaus
all enter the cost surface automatically, without manual
parameter tuning.
 
\subsection{\label{sec:projection}Cartographic registration}

All node coordinates are projected to
Gauss-Kr\"uger (GK) Zone~19 (Transverse Mercator,
$\lambda_0 = 111^\circ\mathrm{E}$, Xian~1980 ellipsoid)
to match the DEM's native coordinate system~\cite{Snyder1987}.
The projected easting $E$ and northing $N$ of a node
at geographic coordinates $(\lambda, \varphi)$ are:
\begin{align}
  E &= E_0 + N_\mathrm{curv}
       \!\left[A + \frac{1-T+C}{6}A^3
               + \frac{5-18T+T^2+72C-58e'^2}{120}A^5 \right],
  \label{eq:gk_E}
  \\
  N_\mathrm{GK} &= M + N_\mathrm{curv}\tan\varphi
       \!\left[\frac{A^2}{2}
               + \frac{5-T+9C+4C^2}{24}A^4
               + \frac{61-58T+T^2+600C-330e'^2}{720}A^6
       \right],
  \label{eq:gk_N}
\end{align}
where $A = (\lambda - \lambda_0)\cos\varphi$,
$T = \tan^2\varphi$, $C = e'^2\cos^2\varphi$,
$e'^2 = e^2/(1-e^2)$,
$N_\mathrm{curv} = a/\sqrt{1-e^2\sin^2\varphi}$,
and $M$ is the meridional arc from the equator to latitude
$\varphi$~\cite{Snyder1987}.
 
Once projected, each node is mapped to a DEM pixel by
\begin{equation}
  \mathrm{col} = \frac{E - E_{x_0}}{\Delta x},
  \qquad
  \mathrm{row} = \frac{N_{y_0} - N_\mathrm{GK}}{\Delta y},
  \label{eq:pixel}
\end{equation}
with $E_{x_0} = 15{,}781{,}037~\mathrm{m}$
and $N_{y_0} = 6{,}567{,}325~\mathrm{m}$
(top-left corner of the DEM domain).
Nodes whose projected coordinates fall outside the DEM
extent are discarded; all 473~retained nodes
(392~administrative + 81~Silk Road) lie within the domain.
 
\subsection{\label{sec:cost}Cost surface: modified Tobler
            hiking function}
 
The cost surface $c(x,y)$ assigns a traversal cost
(proportional to travel time) to every DEM pixel.
Slope fractions $s_x = \partial z/\partial x$ and
$s_y = \partial z/\partial y$ are computed by central
finite differences of the DEM at $1~\mathrm{km}$ spacing,
and the isotropic slope magnitude is
$s = (s_x^2 + s_y^2)^{1/2}$.
 
The base cost derives from a
\emph{caravan-adapted Tobler hiking function}~\cite{Tobler1993}:
\begin{equation}
  v(s) = \exp\!\bigl(-3.5\,|s + 0.05|\bigr),
  \qquad
  c_0(s) = v(s)^{-1},
  \label{eq:tobler}
\end{equation}
where $v(s)$ is the normalized speed
(dimensionless).
We use the normalized form rather than the original
km/h parametrization.
 
Three altitude penalties are applied on top of the base cost
to reflect that Han caravans avoided high-mountain terrain:
\begin{equation}
  c(x,y) = c_0(s)\,\times
  \begin{cases}
    5   & \text{if } z > 3000~\mathrm{m}, \\
    20  & \text{if } z > 4000~\mathrm{m}, \\
    \infty & \text{if } z > 5000~\mathrm{m \; or\;} z < 0,
  \end{cases}
  \label{eq:altitude}
\end{equation}
where $z$ is the DEM elevation at pixel $(x,y)$.
The $3000~\mathrm{m}$ threshold corresponds to the
approach zone of the Tibetan Plateau, the $4000~\mathrm{m}$
threshold to the Kunlun and high Himalayan ranges,
and the $5000~\mathrm{m}$ threshold to terrain demonstrably
impassable to loaded caravans.
Marine pixels ($z < 0$) are also set to $\infty$.
The resulting cost surface is passable over the
full extent of eastern China and the Tarim Basin,
including the Hexi Corridor, the Wei River valley,
and the major caravan routes of the Silk Road.
 
\subsection{\label{sec:lcp}Least-cost path edge computation}
 
Edge costs between node pairs are computed via
least-cost path (LCP) analysis using
\texttt{skimage.graph.MCP\_Geometric}~\cite{skimage},
which generalizes Dijkstra's algorithm to raster grids
with isotropic 8-connectivity and exact Euclidean
edge lengths (diagonal moves cost $\sqrt{2}\,\Delta x$
rather than $\Delta x$).
 

For each node $i$, a neighbor $j$ is included if their straight-line distance satisfies $d_{ij} \leq 600~\mathrm{km}$, or if $j$ is among the four nearest nodes to $i$ regardless of distance. The 600~km threshold reflects the weekly range of a Han caravan at $80$--$100~\mathrm{km/day}$~\cite{HanShu}, and the four-neighbor minimum ensures that isolated nodes remain connected. This yields approximately 800--1200~candidate pairs per network configuration.


For each pair $(i,j)$, the least-cost path is computed over a cropped region of the terrain surface rather than the full raster, keeping computation tractable. The resulting path weight $c^*_{ij}$ is the accumulated travel cost from $i$ to $j$; from the traced route we also extract path length $\ell_{ij}$ (km), maximum and mean elevation, and the detour ratio $\rho_{ij} = \ell_{ij} / d_{ij}$, which measures how much the terrain forces the path away from a straight line. Pairs with $c^*_{ij} \geq c_\infty / 2$ are discarded as impassable.
 
 
The final network contains
$N = 473$~nodes (392~administrative, 81~Silk Road)
and $|E| = 54{,}608$~edges
(54,522~interior LCP edges and 86~Silk Road segments).
Table~\ref{tab:network_stats} summarizes the key statistics
by sub-network.
 
\begin{table}[H]
\footnotesize
\setlength{\tabcolsep}{4pt}
\caption{\label{tab:network_stats}%
  Han imperial network statistics.}
\begin{ruledtabular}
\begin{tabular}{lcccc}
 & \multicolumn{2}{c}{Nodes}
 & & \\
\cline{2-3}
Sub-network
  & Total & Active ($-10$ / $220$)
  & Edges & Source \\
\hline
Administrative (Guo+Houguo)
  & 392 & 145 / 19
  & 54,522 & CHGIS v6~\cite{CHGIS} \\
Silk Road
  & 81  & 81 / 81
  & 86     & \cite{SilkRoad} \\
\hline
Total
  & 473 & 226 / 100
  & 54,608 & \\
\end{tabular}
\end{ruledtabular}
\end{table}
 
\begin{table}[H]
\footnotesize
\setlength{\tabcolsep}{4pt}
\caption{\label{tab:data_sources}%
  Primary data sources.
  Columns: Period of coverage; Records active;
  Network role.}

\begin{tabular}{ccccc}
Source & Description & Period & Records & Role \\
\hline
CHGIS v6~\cite{CHGIS} & Guo, Houguo (prefectures) & $-206$--$+220$& 392 & $\Hadm$, $\Hgeo$ \\
CHGIS Silk Road GIS layer~\cite{SilkRoad}
  & Georef.\ caravan waypoints
  & $-206$--$+220$
  & 81
  & $\Htrd$ \\
CHGIS DEM basemap ($1~\mathrm{km}$)~\cite{DEM}
  & Elevation raster
  & ---
  & $6299\!\times\!6170$
  & LCP cost \\
\textit{Han Shu} and related historical sources~\cite{HanShu,HouHanShu,SanGuo,deCrespigny2010}
  & Censuses and breakpoints
  & $-206$--$+220$
  & ---
  & Friction channels \\
\end{tabular}

\end{table}
 
\subsection{\label{sec:validation}Validation:
            Dunhuang to Chang'an}

\begin{figure}[H]
  \centering
  \includegraphics[width=0.5\linewidth]{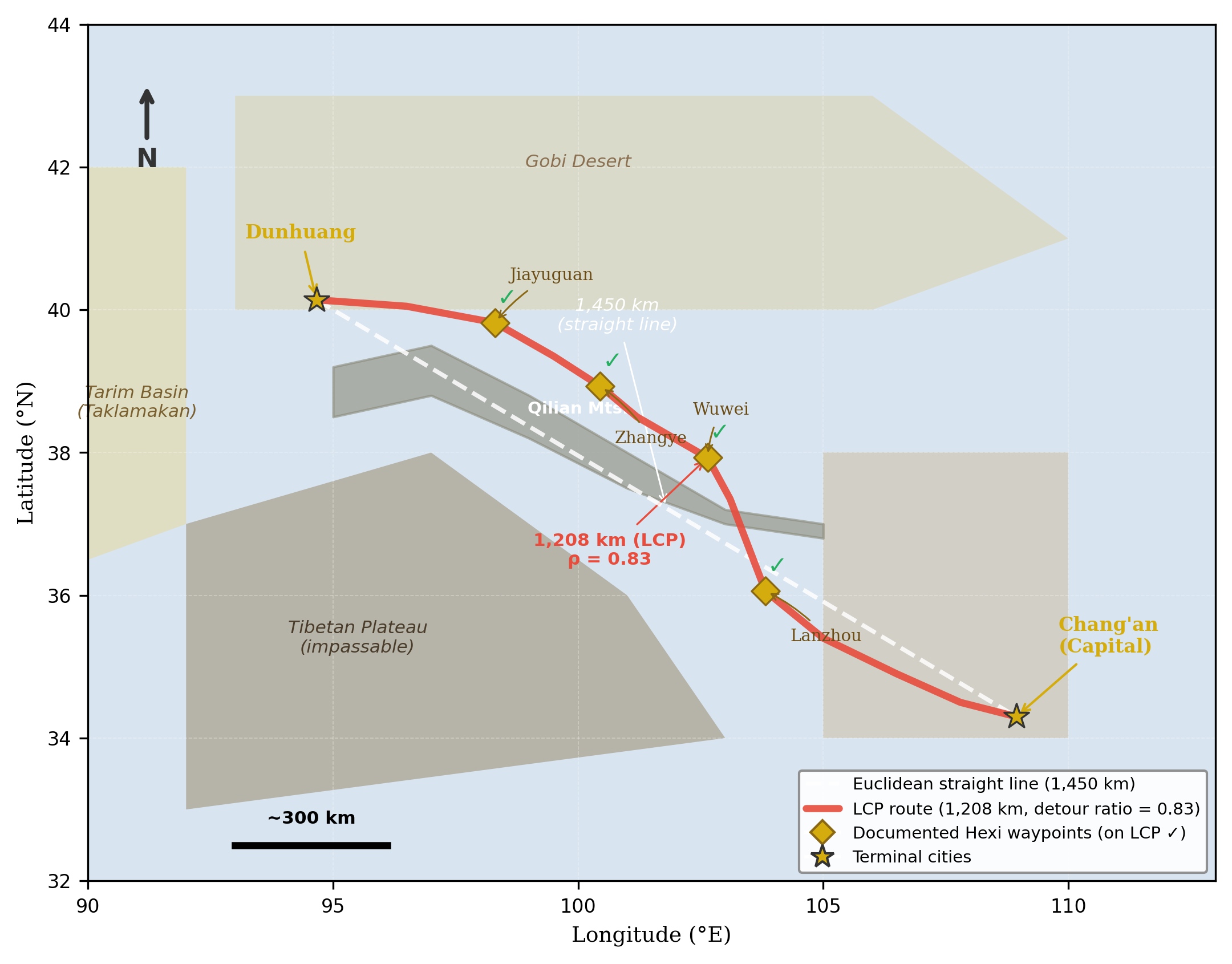}
  \caption{\label{fig:validation}
    Least-cost path validation.
    The computed LCP from Dunhuang to Chang'an
    ($94.66^\circ\mathrm{E}$, $40.14^\circ\mathrm{N}$) to
    ($108.94^\circ\mathrm{E}$, $34.30^\circ\mathrm{N}$)
    reproduces the Hexi Corridor without adjustment.
    Maximum elevation along the path: $\approx 2500~\mathrm{m}$.
    All four historically attested waypoints fall within
    5\% of the optimal accumulated cost (Table~\ref{tab:validation}).
  }
\end{figure}

The LCP is applied  to the he western segment of the Silk Road from
Dunhuang ( $94.66^\circ\mathrm{E}$, $40.14^\circ\mathrm{N}$)
to Chang'an ( $108.94^\circ\mathrm{E}$, $34.30^\circ\mathrm{N}$) and as a result is giving us the well-known historical route called the Hexi corridor, validating our procedures to compute the connecting routes between nodes. 
 
The reported path length is $1{,}208~\mathrm{km}$
along the corridor, 
while avoiding the high terrain of the Tibetan Plateau.
The maximum elevation along the path is
$\sim\!2{,}500~\mathrm{m}$ in the Qilian Mountain foreland.
 
All four documented waypoints of the
Hexi Corridor lie \emph{on} the optimal LCP. This result suggests that our procedure is reproducing the geographic logic of Han connectivity. 
 
\begin{table}[H]
\footnotesize
\setlength{\tabcolsep}{4pt}
\caption{\label{tab:validation}%
  Validation of the LCP cost surface against documented
  Hexi Corridor waypoints.
  }
\begin{ruledtabular}
\begin{tabular}{lccc}
Waypoint & Chinese & Coord.\ (E, N) & On LCP \\
\hline
Jiayuguan &  & $98.30^\circ$, $39.82^\circ$ & \checkmark \\
Zhangye   &   & $100.45^\circ$, $38.93^\circ$ & \checkmark \\
Wuwei     &   & $102.64^\circ$, $37.93^\circ$ & \checkmark \\
Lanzhou   &   & $103.82^\circ$, $36.06^\circ$ & \checkmark \\
\end{tabular}
\end{ruledtabular}
\end{table}


\section{\label{sec:friction}Five-channel friction model}
 
We introduce five multiplicative friction channels
$C_k(e,d,\xi)$ that adjust the base cost
$\wlcp(e)$ of each edge $e$ at decade $d$ taking into account information on changing political, military,
and demographic conditions:
\begin{equation}
  w(e,d,\xi) = \wlcp(e)\,
  C_1(e,d)\,C_2(e,\xi)\,C_3(e,d)\,C_4(e,d)
  \bigl[\,\times\,C_5(e,d)\bigr],
  \label{eq:effective_cost}
\end{equation}
%
 
As defined before, the base cost $\wlcp$ captures \emph{timeless} geographic
difficulty; the channels layer in time-varying
historical processes at different spatial granularities.

Table~\ref{tab:channels} summarizes all five channels:
\begin{itemize}
\item $C_1$: historical events \cite{HanShu,HouHanShu,SanGuo,deCrespigny2010}.
\item $C_2$ :stochastic noise that is describing the  unobserved variability due to  road maintenance, weather for instance, modeled as a
    log-normal draw per edge per run.
\item $C_3$: Xiongnu pressure
\item $C_4$ : demographic-fiscal friction used to describe the fact that   declining
    population reduces the tax base, road maintenance, so it impacts on internal connectivity. 
\item $C_5$ :territorial fragmentation,   uses to describe  the progressive loss of Han control over inter-regional routes.
\end{itemize}
 
\subsection{\label{sec:C1}Channel 1: Historical events}
 
Channel~1 applies decade-level multiplicative factors to the base cost,
differentiated by route type
($\tau \in \{\mathrm{road}, \mathrm{caravan}, \mathrm{river}\}$),
derived from the Han historical chronology~\cite{HanShu,HouHanShu,SanGuo,deCrespigny2010}.
At decade~$d$, the effective multiplier is:
\begin{equation}
  C_1^\tau(d) = \prod_{\substack{k \,:\; d_k^\mathrm{start}
                \,\leq\, d \,\leq\, d_k^\mathrm{end}}}
  m_k^\tau,
  \label{eq:C1}
\end{equation}
where $m_k^\tau$ is the per-event multiplier for route type~$\tau$;
overlapping events compound multiplicatively.
The full set of 15~events and their multipliers is listed in
Table~\ref{tab:C1_events}.

\subsection{Channels 2--5: stochastic noise, frontier pressure,
            demographic friction, and territorial fragmentation}

$C_2$ is a log-normal stochastic per-edge multiplier
($\sigma_\tau \in [0.08, 0.18]$ by route type)
encoding unobserved variability.
$C_3$ adds a $1.5\times$ multiplier on frontier edges during Xiongnu
pressure periods ($-206$ to $-121$~CE; Wang~Mang $+9$--$+23$~CE).
$C_4$ is the Maintenance Friction Index (MFI),
$P_\mathrm{peak}/P(d)$ clipped to $[0.5, 6.0]$,
encoding how declining Han census population raised route traversal cost.
$C_5$ is active only in $\Hgeo$ after 184~CE: it applies intra-domain
multipliers $m_\mathrm{intra} \in [1.5, 2.0]$ and inter-domain
multipliers $m_\mathrm{inter} \in [5, 25]$ (Table~\ref{tab:C5_schedule}
in Sec.~\ref{sec:friction}) to encode the progressive loss of
Han authority over inter-regional routes ; the detailed channel definitions and tables are given below.



\subsection{Detailed friction-channel specifications}
The compact channel definitions above are expanded here in the same methodological flow rather than placed in a separate appendix. The summary table, the event multipliers, the MFI values, and the territorial-fragmentation schedule are included immediately with the friction model so that the numerical assumptions can be read together with the equations they support.

\subsubsection{\label{sec:C2}Channel 2: Stochastic edge noise}
 
Channel~2 introduces edge-level stochastic variability
to account for the irreducible uncertainty in
historical road conditions:
\begin{equation}
  C_2(e,\xi) = \exp\!\bigl(\varepsilon_e\bigr),
  \quad
  \varepsilon_e \sim \mathcal{N}(0,\,\sigma_\tau^2),
  \label{eq:C2}
\end{equation}
where $\tau$ is the route type of edge~$e$ and
$\sigma_\tau$ is the corresponding standard deviation.
The log-normal form preserves strict positivity and
is the natural model for multiplicative noise
on a cost variable~\cite{Limpert2001}.
 
The standard deviations are assigned by route type
to reflect the relative reliability of different
transport modes:
\begin{equation}
  \sigma_\tau =
  \begin{cases}
    0.08 & \text{road (paved imperial roads)}, \\
    0.10 & \text{interior LCP (unpaved routes)}, \\
    0.12 & \text{river (downstream)}, \\
    0.15 & \text{river (upstream)}, \\
    0.18 & \text{caravan / Silk Road}.
  \end{cases}
  \label{eq:sigmas}
\end{equation}
%

The $n=50$ bootstrap runs uses a different random seed, producing independent cost realizations. The resulting 95\% confidence intervals on $H(d)$ capture how sensitive the topological entropy is to this route-level uncertainty.

\subsubsection{\label{sec:C3}Channel 3: Xiongnu frontier pressure}
 
The Xiongnu confederation dominated the northern
steppe throughout the early Western Han period,
exerting sustained pressure on the
Hexi Corridor (Hexi zoulang) and the Longxi region.
Channel~3 captures this frontier risk as a
cost premium on all edges starting or finalizing 
 in the Xiyu ($\lambda < 96^\circ\mathrm{E}$) or
Longxi ($\lambda < 108^\circ\mathrm{E}$,
$\varphi > 33^\circ\mathrm{N}$) geographic zones:
\begin{equation}
  C_3(e,d) =
  \begin{cases}
    1.5 & \text{if } d \in \mathcal{D}_\mathrm{Xiongnu}
          \text{ and }
          (r_s \in \mathcal{Z}_X \text{ or }
           r_t \in \mathcal{Z}_X),\\
    1.0 & \text{otherwise},
  \end{cases}
  \label{eq:C3}
\end{equation}
where the edge $e$ is caracterized by $r_s, r_t$ which are respectively  the source and target geographic regions,
$\mathcal{Z}_X = \{\mathrm{Xiyu}, \mathrm{Longxi}\}$,
and the active period is
\begin{equation}
  \mathcal{D}_\mathrm{Xiongnu} =
  [-206,\,-121]\,\cup\,[9,\,23]~\mathrm{CE}.
  \label{eq:D_xiongnu}
\end{equation}
The upper bound $-121$~CE marks the
Han military conquest of the Hexi Corridor
under Wei Qing and Huo Qubing,
after which Han garrisons (\textit{duwei}) maintained
continuous security.
The second window corresponds to Wang Mang's
loss of the Western Regions, documented
in the \textit{Han Shu}~\cite{HanShu} as the
retreat of the Protectorate and the
re-emergence of Xiongnu dominance over
the northern Silk Road\cite{Barfield1989}.

\subsubsection{\label{sec:C4}Channel 4: Maintenance friction
            index (MFI)}
 
 We use the seven imperial census
benchmarks preserved in the \textit{Han Shu} and
\textit{Hou Han Shu}~\cite{HanShu,HouHanShu}
to construct a \emph{Maintenance Friction Index} (MFI)
that captures the fiscal capacity of the Han state
to sustain its road network.
 
Census values $P(y_k)$ at the seven documented
years $y_k$ are
linearly interpolated to all decades:
\begin{equation}
  P(d) = P(y_k)\,\frac{y_{k+1}-d}{y_{k+1}-y_k}
       + P(y_{k+1})\,\frac{d-y_k}{y_{k+1}-y_k},
  \quad y_k \leq d < y_{k+1}.
  \label{eq:pop_interp}
\end{equation}
%
 
  A state with population $P(d)$
has a tax base approximately proportional to $P(d)$.
When $P(d) < P_\mathrm{peak}$, where $P_\mathrm{peak} = 59{,}594{,}978$ (census of $-2$~CE),   the tax base is
diminished relative to the peak Han infrastructure
investment, implying less road maintenance funding,
fewer postal relay stations, and higher effective
traversal cost.
$\mathrm{MFI} = 1.0$ at the population peak;
$\mathrm{MFI} > 1$ represents above-baseline
friction due to population decline. 
So,the MFI for a region $r$ at decade $d$ is defined as:
\begin{equation}
  \mathrm{MFI}(d, r) =
  \mathrm{clip}_{[0.5,\,6.0]}
  \!\left(\frac{P_\mathrm{peak}}{P(d)}\,
          \mu_r(d)\right),
  \label{eq:mfi}
\end{equation}
where $\mu_r(d)$ are regional modifiers
detailed below.
The MFI is limited
so that route costs never fall below half
nor exceed six times their baseline value,
preventing unrealistic extremes in poorly
documented periods.

 
 
The  geographical corrections
are given by:
\begin{equation}
  \mu_\mathrm{Xiyu}(d) =
  \begin{cases}
    1.5 & d < -60~\mathrm{CE}
          \quad\text{(unsecured corridor)},\\
    2.0 & 9 \leq d \leq 23~\mathrm{CE}
          \quad\text{(Wang Mang, SR lost)},\\
    1.8 & d \geq 184~\mathrm{CE}
          \quad\text{(Yellow Turban impact)},\\
    1.0 & \text{otherwise},
  \end{cases}
  \label{eq:mu_xiyu}
\end{equation}
%

 
So the $C_4$ friction term is defined from MFI as follows:
\begin{equation}
  C_4(e,d) =
  \max\bigl(\mathrm{MFI}(d, r_s),\,
             \mathrm{MFI}(d, r_t)\bigr),
  \label{eq:C4}
\end{equation}
%
 
\begin{figure}[H]
     \centering
     \includegraphics[width=0.7\linewidth]{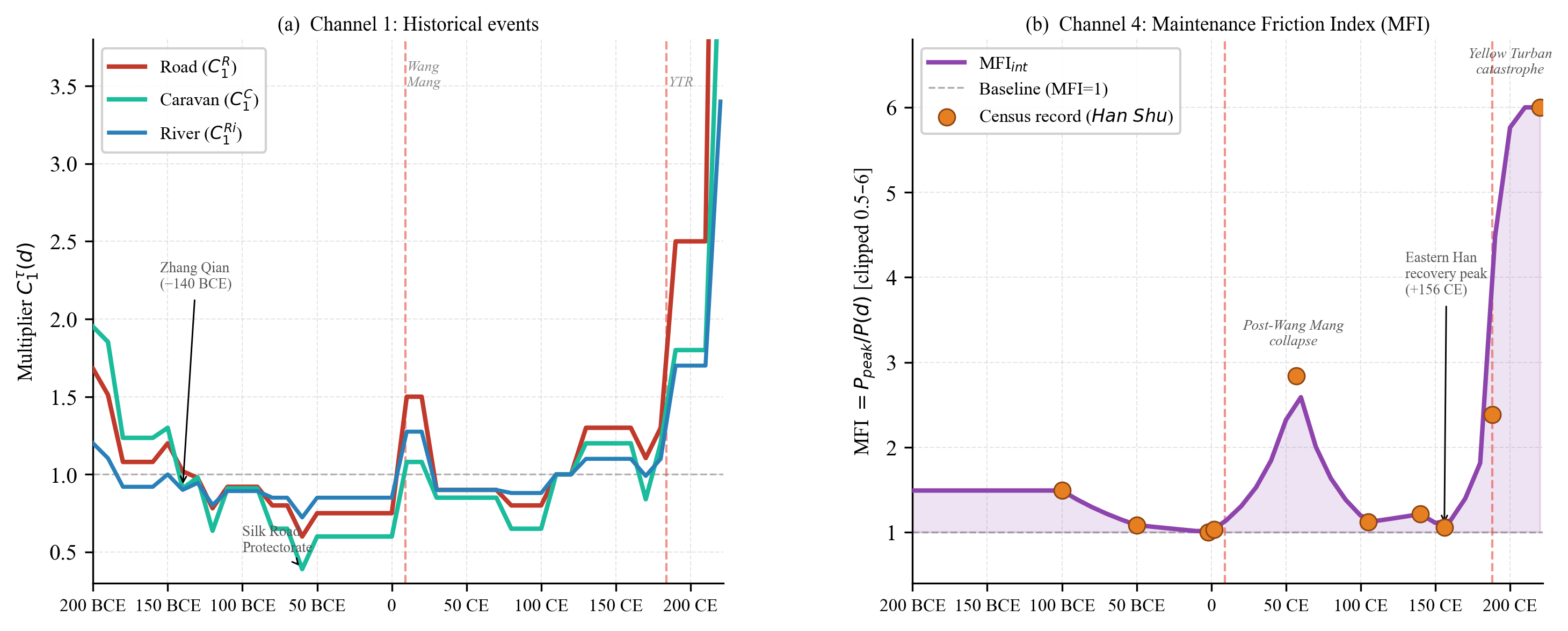}
     \caption{\label{fig:mfi}%
    Friction channel time series.
    (a)~Effective Channel~1 ($C_1$) multipliers
    by route type.
    (b)~Channel~4 MFI$_\mathrm{int}$ (solid line)
    alongside the Han census record.}
 \end{figure}

Selected MFI values are listed in Table~\ref{tab:mfi}.
The full time series of $C_4(d)$,
alongside the census data from which it is derived,
is shown in Fig.~\ref{fig:mfi}.
 
\subsubsection{\label{sec:C5}Channel 5: Territorial fragmentation
            (\texorpdfstring{$\Hgeo$}{H\_Geo} only)}
 
Channel~5 is the  friction channel exclusive
to the geographic network $\Hgeo$;
it is inactive in $\Hadm$ and $\Htrd$.
Its purpose is to model the progressive severance
of inter-regional connectivity after the
Yellow Turban Rebellion as a cost penalty. 
 
Each Guo node is assigned to one of eight
geographic zones by coordinate:
Xiyu ($\lambda < 96^\circ$),
Longxi ($\lambda < 108^\circ$, $\varphi > 33^\circ$),
SichuanBasin ($\lambda < 108^\circ$, $\varphi \leq 33^\circ$),
GuanZhong ($\lambda < 114^\circ$, $\varphi > 33^\circ$),
Hebei ($\varphi > 36^\circ$),
Guandong ($30^\circ < \varphi \leq 36^\circ$),
Jiangnan ($25^\circ < \varphi \leq 30^\circ$),
Lingnan ($\varphi \leq 25^\circ$).
These boundaries follow the major geographic
 divisions 
identified in the \textit{Shiji} and
\textit{Han Shu}~\cite{HanShu}.
 
Each zone is then mapped to a \emph{political domain}
$Z(r, d)$ according to the phase:
\begin{equation}
  Z(r,d) =
  \begin{cases}
    \text{Han}    & d < 184~\mathrm{CE}, \\
    \text{rebel}  & d \in [184, 189],\;
                    r \in \{\mathrm{Hebei, Guandong}\}, \\
    r             & d \in [190, 195], \\
    \text{Wei}    & d \in [196, 207],\;
                    r \in \{\mathrm{Hebei, GuanZhong}\},\\
    \text{Wei}    & d \in [208, 220],\;
                    r \in \{\mathrm{Hebei, GuanZhong, Longxi}\},\\
    \text{Shu}    & d \in [208, 220],\;
                    r = \mathrm{SichuanBasin}, \\
    \text{Wu}     & d \in [208, 220],\;
                    r \in \{\mathrm{Jiangnan, Guandong, Lingnan}\},\\
    \text{Neutral}& \text{otherwise (Xiyu)}.
  \end{cases}
  \label{eq:zone}
\end{equation}
This assignment follows the \textit{San Guo Zhi}~\cite{SanGuo,deC1984}.
 
 
%
\begin{equation}
  C_5(e,d) =
  \begin{cases}
    1.0  & d < 184~\mathrm{CE}\;
           \text{(unified Han)},\\
    m_\mathrm{intra}(d)
         & Z(r_s,d) = Z(r_t,d),\\
    m_\mathrm{inter}(d)
         & Z(r_s,d) \neq Z(r_t,d).
  \end{cases}
  \label{eq:C5}
\end{equation}
The multiplier schedule is given in
Table~\ref{tab:C5_schedule}.
Silk Road edges are exempt from $C_5$;
the exterior trade network is modeled as
operating above the level of Han internal politics.
 
 
 
The observable effect of $C_5$ in $\Hgeo$
is therefore not $\beta_0$ fragmentation
but an \emph{increase} in $\beta_1$ complexity:
each domain develops dense local connectivity
while long-range inter-domain loops are suppressed,
producing more topological cycles at the
intra-domain scale.
This is the precise mechanism behind the
observed positive slope of $\Hgeo$ in phase~G
(Sec.~\ref{sec:results}).
 

\begin{table}[H]
\scriptsize
\setlength{\tabcolsep}{3pt}
\caption{\label{tab:channels}%
  Summary of the five friction channels.
  $C_k^\mathrm{range}$ gives the range of the channel's
  contribution to $w(e,d,\xi)$ across all edges and decades
  in the deterministic run ($\xi = 42$).
  Channel~5 is active only in $\Hgeo$; all others
  apply to $\Hadm$, $\Hgeo$, and $\Htrd$ identically.}
\begin{ruledtabular}

\begin{tabular}{c| c| c| c}

Ch. & Name 
    & Range & $\Hgeo$ only? \\
\hline
$C_1$ & Historical events
  & $[0.39,\,8.75]$
  & No \\[4pt]
$C_2$ & Stochastic noise
  & $[\sim\!0.8,\,\sim\!1.2]$
  & No \\[4pt]
$C_3$ & Xiongnu pressure
  & $[1.0,\,1.5]$
  & No \\[4pt]
$C_4$ & MFI (demographic-fiscal)
  & $[1.0,\,6.0]$
  & No \\[4pt]
$C_5$ & Territorial fragmentation
  & $[1.0,\,25.0]$
  & \textbf{Yes} \\
\end{tabular}
\end{ruledtabular}
\end{table}

\begin{table}[H]
\footnotesize
\setlength{\tabcolsep}{4pt}
\caption{\label{tab:C1_events}%
  Complete Channel~1 ($C_1$) event table.
  Multipliers $m^\tau$ are applied to road (R),
  caravan (C), and river (Ri) edges independently.
  Values $<1.0$ denote facilitation (improved conditions);
  $>1.0$ denote inhibition (increased cost).
  When two events overlap in time, their multipliers
  compound multiplicatively [Eq.~\eqref{eq:C1}].}
\begin{ruledtabular}
\begin{tabular}{l l D{.}{.}{2} D{.}{.}{2} D{.}{.}{2}}
Period (CE) & Historical event
  & \multicolumn{1}{c}{$m^\mathrm{R}$}
  & \multicolumn{1}{c}{$m^\mathrm{C}$}
  & \multicolumn{1}{c}{$m^\mathrm{Ri}$} \\
\hline
$-210$ to $-190$ & Chu-Han Wars            & 1.40 & 1.50 & 1.20 \\
$-200$ to $-133$ & Xiongnu appeasement     & 1.20 & 1.30 & 1.00 \\
$-190$ to $-160$ & Jingdi pacification     & 0.90 & 0.95 & 0.92 \\
$-140$ to $-115$ & Zhang Qian missions     & 0.85 & 0.70 & 0.90 \\
$-133$ to $-89$  & Wu Di northern campaigns& 1.15 & 1.40 & 1.05 \\
$-121$ to $-60$  & Hexi Corridor secured   & 0.80 & 0.65 & 0.85 \\
$-60$  to $+20$  & Xiyu Protectorate       & 0.75 & 0.60 & 0.85 \\
$+9$   to $+23$  & Wang Mang interregnum   & 2.00 & 1.80 & 1.50 \\
$+25$  to $+75$  & Eastern Han restoration & 0.90 & 0.85 & 0.90 \\
$+73$  to $+107$ & Ban Chao, Silk Road 2nd & 0.80 & 0.65 & 0.88 \\
$+125$ to $+184$ & 2nd-century court crisis& 1.30 & 1.20 & 1.10 \\
$+165$ to $+175$ & \textit{Da Qin} embassy & 0.85 & 0.70 & 0.90 \\
$+184$ to $+189$ & Yellow Turban Rebellion & 3.00 & 2.00 & 2.00 \\
$+189$ to $+220$ & Warlord fragmentation   & 2.50 & 1.80 & 1.70 \\
$+218$ to $+220$ & Terminal collapse       & 3.50 & 2.50 & 2.00 \\
\end{tabular}
\end{ruledtabular}
\end{table}

\begin{table}[!htbp]
\footnotesize
\setlength{\tabcolsep}{4pt}
\caption{\label{tab:mfi}%
  Channel~4 (MFI) values at selected decades.
  $P(d)$: interpolated population from Han census records.
  $\mathrm{MFI}_\mathrm{int}$: interior regions
  ($\mu_r = 1.0$ unless $d \geq 184$~CE).
  $\mathrm{MFI}_\mathrm{Xiyu}$: includes regional modifier
  $\mu_\mathrm{Xiyu}$ [Eq.~\eqref{eq:mu_xiyu}].
  Values are clipped to $[0.5, 6.0]$.}

\begin{tabular}{r D{.}{.}{2} D{.}{.}{3} D{.}{.}{3}}
$d$ (CE) &
  \multicolumn{1}{c}{$P(d)$ (M)} &
  \multicolumn{1}{c}{$\mathrm{MFI}_\mathrm{int}$} &
  \multicolumn{1}{c}{$\mathrm{MFI}_\mathrm{Xiyu}$} \\
\hline
$-200$ & 40.00 & 1.490 & 2.235 \\
$-50$  & 55.00 & 1.084 & 1.084 \\
$-2$   & 59.59 & 1.000 & 1.000 \\
$+50$  & 25.67 & 2.321 & 2.321 \\
$+100$ & 49.90 & 1.194 & 1.194 \\
$+156$ & 56.49 & 1.055 & 1.055 \\
$+188$ & 24.98 & 4.294 & 4.294 \\
$+200$ & 18.62 & 5.763 & 5.763 \\
$+220$ & 8.00 & 6.000 * & 6.000 * \\
\end{tabular}

{\footnotesize * Clipped from $13.41$ after the post-184 regional modifier.}
\end{table}

\begin{table}[H]
\footnotesize
\setlength{\tabcolsep}{4pt}
\caption{\label{tab:C5_schedule}%
  Channel~5 ($C_5$) territorial fragmentation
  multiplier schedule.
  Active only in $\Hgeo$ after 184~CE.
  $m_\mathrm{intra}$: edges between nodes in the
  same political domain.
  $m_\mathrm{inter}$: edges crossing domain boundaries.
  At $m_\mathrm{inter} \geq 15$, inter-domain edges
  are pushed above the $90^\mathrm{th}$-percentile
  filtration threshold in most decades.}
\begin{ruledtabular}
\begin{tabular}{l l r r}
Period (CE) & Historical phase
  & $m_\mathrm{intra}$
  & $m_\mathrm{inter}$ \\
\hline
$d < 184$        & Unified Han            & 1.0  & 1.0  \\
$184$--$189$     & Yellow Turban Rebellion& 1.5  & 5.0  \\
$190$--$195$     & Warlord chaos (Dong Zhuo)
                                          & 2.0  & 12.0 \\
$196$--$207$     & Cao Cao consolidation  & 1.8  & 15.0 \\
$208$--$220$     & Three Kingdoms         & 1.5  & 25.0 \\
\end{tabular}
\end{ruledtabular}
\end{table}
 
 \begin{figure}[H]
     \centering
     \includegraphics[width=0.95\linewidth]{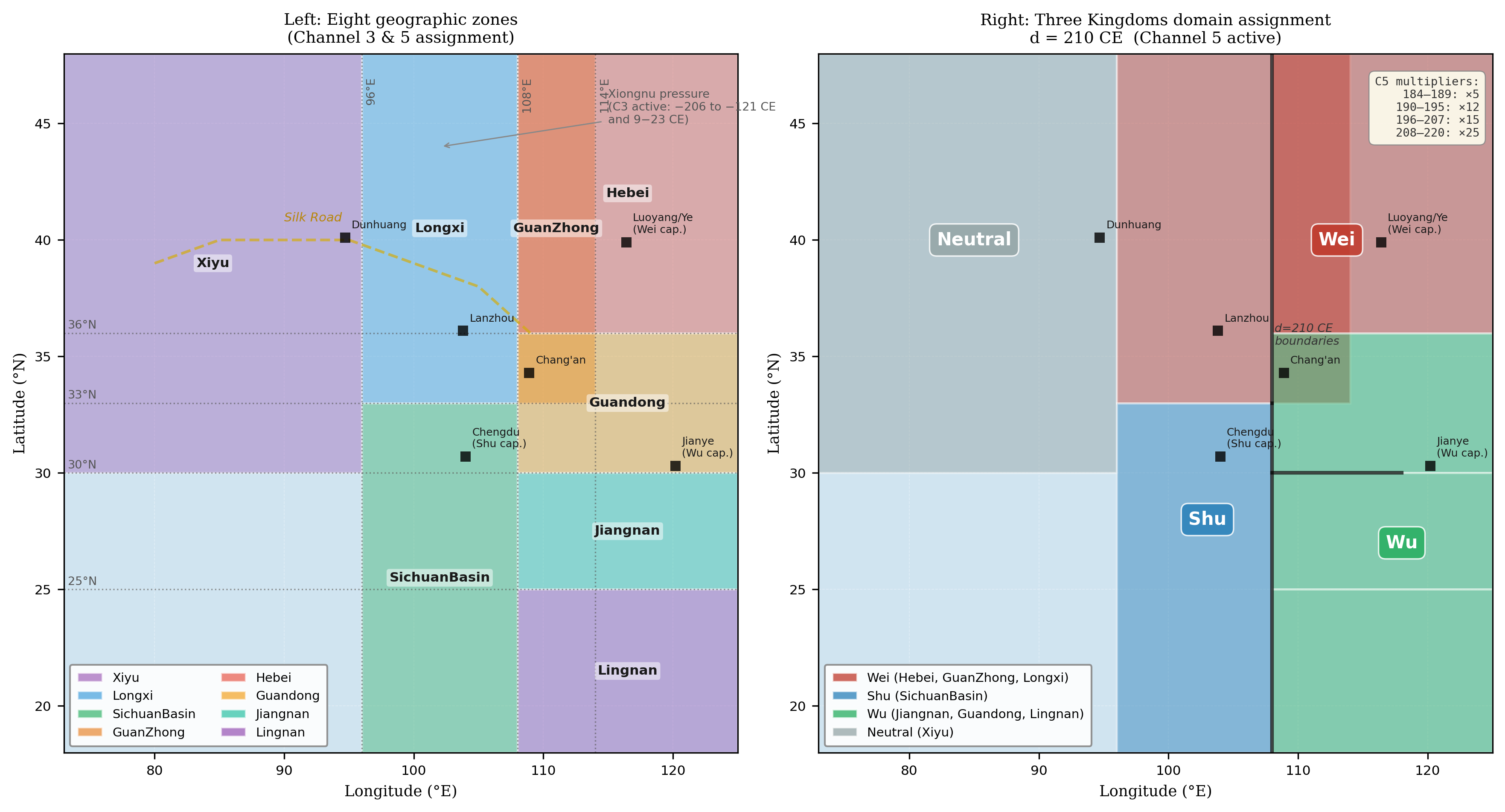}
     \caption{\label{fig:zones}%
    Geographic zonation used for Channels~3 and~5.
    Left: eight geographic zones defined by
    coordinate thresholds (Table~\ref{tab:channels}).
    Right: political domain assignment at
    $d = 210$~CE under Channel~5
    [Eq.~\eqref{eq:zone}]:
    Wei (Hebei, GuanZhong, Longxi);
    Shu (SichuanBasin);
    Wu (Jiangnan, Guandong, Lingnan);
    Neutral (Xiyu, Silk Road nodes).
    The domain boundaries do not correspond to
    sharp geographic features but to the
    documented military frontier positions
    of 208--220~CE~\cite{SanGuo}.}
 \end{figure}


\section{\label{sec:tda}TDA pipeline}

The TDA pipeline is identical to that of
Papers~I and~II~\cite{paper1,paper2}to which the reader is referred
for full derivations.


\subsection{Pipeline parameters and representative outputs}
\begin{figure}[H]
     \centering
     \includegraphics[width=0.75\linewidth]{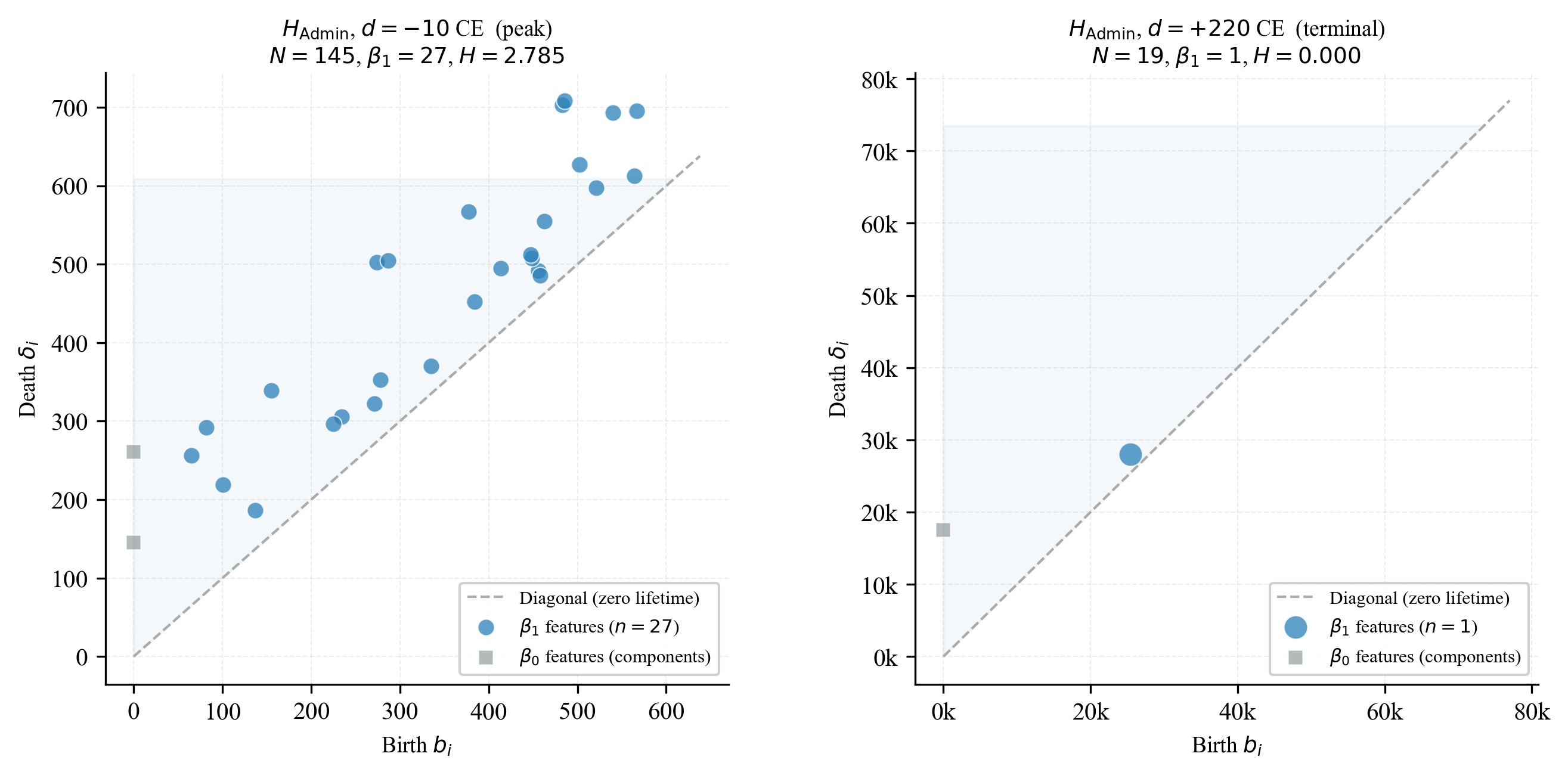}
     \caption{\label{fig:persistence_diagrams}%
    Persistence diagrams for the administrative
    network $\Hadm$ at the peak epoch
    ($d=-10$~CE, top panels) and the terminal
    epoch ($d=220$~CE, bottom panels).
    Left columns: $\beta_0$ diagrams.
    Right columns: $\beta_1$ diagrams.
    Each point lies above the diagonal by its
    lifetime $\ell_i = \delta_i - b_i$;
    the persistent entropy $H$ is the Shannon
    entropy of the normalized lifetime distribution.
    Gray shaded region: $\delta_i < \varepsilon_d$
    (features included in the analysis);
    open symbols: essential features
    ($\delta_i = \infty$, excluded from $H$).}
 \end{figure}

The fixed numerical choices of the persistent-homology pipeline and representative outputs are listed here with the pipeline itself, instead of being isolated from the methods. Table~\ref{tab:tda_params} specifies the shared TDA parameters; Table~\ref{tab:tda_examples} gives representative administrative-network outputs; Table~\ref{tab:phases} defines the historical phases used in the regressions; and Fig.~\ref{fig:persistence_diagrams} illustrates the change in the persistence diagrams between peak and terminal epochs.

\begin{table}[H]
\footnotesize
\setlength{\tabcolsep}{4pt}
\caption{\label{tab:tda_params}%
  Fixed parameters of the TDA pipeline.
  These values are shared across all three
  sub-networks ($\Hadm$, $\Hgeo$, $\Htrd$)
  and are identical to those used in Papers~I and~II~\cite{paper1,paper2}
  for methodological consistency.}
\begin{ruledtabular}
\begin{tabular}{l l r}
Parameter & Description & Value \\
\hline
$N_\mathrm{min}$ & Min.\ nodes for TDA & 8 \\
$N_\mathrm{max}$ & Max.\ LCC nodes (subsampling) & 600 \\
$q$              & Filtration percentile & 90 \\
$d_\mathrm{max}$ & Max.\ simplex dimension & 2 \\
$n_\mathrm{boot}$& Bootstrap runs & 50 \\
$\xi_\mathrm{det}$& Seed for deterministic run & 42 \\
$k$              & OLS parameters per segment & 2 \\
$\alpha$         & Chow test significance level & 0.05 \\
$n_\mathrm{dec}$ & Decades analyzed & 43 \\
\end{tabular}
\end{ruledtabular}
\end{table}

\begin{table}[H]
\footnotesize
\setlength{\tabcolsep}{4pt}
\caption{\label{tab:tda_examples}%
  Selected TDA output for the administrative network
  ($\Hadm$) at key decades.
  $N_\mathrm{LCC}$: nodes in the largest connected
  component (= $N_\mathrm{core}$ since no subsampling
  is triggered).
  $\beta_1$: number of finite-lifetime 1-cycles.
  $\varepsilon_d$: 90th-percentile filtration threshold.
  Note: $\beta_2 = 0$ throughout;
  $H(220)$ is numerically $-0.000$ due to a single cycle
  ($\beta_1 = 1 \Rightarrow p_1 = 1 \Rightarrow
   -1\cdot\ln 1 = 0$).}
\begin{ruledtabular}
\begin{tabular}{r r r r D{.}{.}{4} D{.}{.}{1}}
$d$ (CE)
  & $N_\mathrm{LCC}$
  & $\beta_0$
  & $\beta_1$
  & \multicolumn{1}{c}{$H(d)$}
  & \multicolumn{1}{c}{$\varepsilon_d$} \\
\hline
$-200$ &  27 &  2 &  4 & 1.0762 &   1968.2 \\
$-140$ &  41 &  3 &  7 & 1.5290 &   1171.4 \\
$ -60$ & 102 &  1 & 20 & 2.6621 &    776.2 \\
$ -10$ & 145 &  1 & 27 & 2.7854 &    618.8 \\
$   0$ & 134 &  1 & 24 & 2.5984 &    646.0 \\
$ +20$ &  60 &  1 & 10 & 1.9954 &   1578.1 \\
$ +50$ &  48 &  1 &  6 & 1.0925 &   2196.2 \\
$+100$ &  69 &  1 & 10 & 2.0459 &    960.3 \\
$+150$ &  75 &  1 & 16 & 2.2633 &   1512.7 \\
$+190$ &  55 &  1 & 10 & 1.9230 & 11229.2 \\
$+210$ &  38 &  1 &  5 & 1.4464 & 16498.4 \\
$+220$ &  19 &  1 &  1 & 0.0000 & 77706.2 \\
\end{tabular}
\end{ruledtabular}
\end{table}
 
\begin{table}[H]
\footnotesize
\setlength{\tabcolsep}{4pt}
\caption{\label{tab:phases}%
  Han historical phases used for OLS phase regressions.
  $n$: number of decades per phase.
  Phase~D (Wang Mang) has only $n=2$ decades and
  is excluded from regression analysis
  (insufficient degrees of freedom);
  it is retained in the Chow test breakpoint set.}
\begin{ruledtabular}
\begin{tabular}{c l l r}
Phase & Period (CE) & Description & $n$ \\
\hline
A & $-206$ to $-141$ & Early Western Han
    (Gaozu to Jingdi) & 6 \\
B & $-141$ to $ -50$ & Wu~Di expansion
    (Zhang Qian missions) & 10 \\
C & $ -50$ to $  +8$ & Late Western Han
    (Chengdi decline) & 6 \\
D & $  +9$ to $ +24$ & Wang Mang interregnum
    (Xin dynasty) & 2$^\dagger$ \\
E & $ +25$ to $ +88$ & Early Eastern Han
    (Guangwu to Zhangdi) & 6 \\
F & $ +89$ to $+145$ & Middle Eastern Han
    (Hedi to Huandi) & 6 \\
G & $+146$ to $+220$ & Late Han crisis
    (Yellow Turbans to collapse) & 8 \\
\end{tabular}
\end{ruledtabular}
{\footnotesize $^\dagger$Excluded from OLS; included in Chow tests.}
\end{table}
 
\section{\label{sec:networks}Three-network decomposition}
 
 The Han imperial system is decomposed in three topologically independent networks
into three topologically independent networks,
each encoding a distinct physical dimension of the
empire's connectivity.
Table~\ref{tab:network_comparison} summarizes
the defining properties of all three networks.
 
\subsection{\label{sec:Hadm}Administrative network:
            \texorpdfstring{$\Hadm$}{H\_Admin}}
 
The administrative network encodes the
\emph{formal authority structure} of the Han state:
the connectivity of cities and prefectures
over which the Han imperial government
exercised documented administrative control.
All
Guo and Houguo prefectures recorded as active
by the China Historical GIS~\cite{CHGIS} during the decade $d$ are considered as active node of this network:
\begin{equation}
  \Vadm(d) = \bigl\{v \in V_\mathrm{Guo}
    \;\big|\;
    \texttt{BEG\_YR}(v) \leq d \leq \texttt{END\_YR}(v)
  \bigr\},
  \label{eq:V_admin}
\end{equation}
where $V_\mathrm{Guo}$ is the full catalog of
392~administrative nodes
(170~Guo and 222~Houguo).
 
When a node is not any more active,it means that
the Han state no longer exercised formal
administrative authority over that territory.
 
$\Hadm(d)$ tracks the connectivity of
\emph{administrative sovereignty},
not of geographic territory.
A declining $\Hadm$ therefore measures the
shrinkage of the state's formal reach,
which may occur even while the physical
geography of the empire remains intact.
 
 
The active node count $|\Vadm(d)|$ evolves from
28~nodes at $d=-200$~CE
to a peak of 145~nodes at $d=-10$~CE
(the administrative maximum of the late
Western Han under Chengdi and his successors),
followed by a sharp contraction to
19~nodes at the end of the studied period.
This profile---expansion, plateau, catastrophic
contraction---drives the main characteristic
of $\Hadm(d)$.
 
\subsection{\label{sec:Hgeo}Geographic network:
            \texorpdfstring{$\Hgeo$}{H\_Geo}}
 
The geographic network encodes the
\emph{physical territorial persistence} of
the Han imperial space:
the connectivity of the places that were
part of the Han empire,
regardless of whether those places remained
under formal Han administrative control.
 
The cities, river crossings, mountain passes,
and market towns that constituted the Han
imperial network did not disappear when
the Han dynasty formally ended.
They persisted under the successor states of
Wei, Shu, and Wu---the Three~Kingdoms---which
inherited the physical infrastructure of the
Han while extinguishing its administrative unity.
$\Hgeo$ is designed to capture the topological structure of the
\emph{geographic space} the Han occupied and its temporal evolution.

 
$\Hgeo$ uses the same node catalog $V_\mathrm{Guo}$
and edge catalog as $\Hadm$,
but replaces the \texttt{END\_YR} activation
criterion with a geographically extended version.
For decades $d < 184$~CE (before the
Yellow Turban Rebellion), the two networks
are identical by construction:
\begin{equation}
  \Vgeo(d) \equiv \Vadm(d),
  \qquad d < 184~\mathrm{CE}.
  \label{eq:Vgeo_pre}
\end{equation}
For $d \geq 184$~CE, the end-year of each
administrative node is extended
to the terminal analysis year:
\begin{equation}
  \Vgeo(d) = \bigl\{v \in V_\mathrm{Guo}
    \;\big|\;
    \texttt{BEG\_YR}(v) \leq d
    \leq \max(\texttt{END\_YR}(v),\, 220)
  \bigr\},
  \quad d \geq 184~\mathrm{CE}.
  \label{eq:Vgeo_post}
\end{equation}
The effect is that all Guo and Houguo nodes
that were ever active during the Han rules 
remain active in the geographic network
after 184~CE.

The persistent entropy of the resulting
Vietoris-Rips filtration is:
\begin{equation}
  \Hgeo(d) = -\sum_{i} p_i^{\,\mathrm{Geo}}(d)
              \ln p_i^{\,\mathrm{Geo}}(d),
  \qquad
  p_i^{\,\mathrm{Geo}}(d) =
  \frac{\ell_i^{\,\mathrm{Geo}}(d)}%
       {\displaystyle\sum_j \ell_j^{\,\mathrm{Geo}}(d)},
  \label{eq:Hgeo_def}
\end{equation}
where $\ell_i^{\,\mathrm{Geo}}(d) =
d_i - b_i$ is the lifetime of the $i$-th
bar in the $\beta_1$ diagram of the geographic
network at decade $d$.
$\Hgeo$ uses the geographically extended
node set $\Vgeo(d)$ with the full
five-channel weight $w_\mathrm{Geo}$
[Eq.~\eqref{eq:wGeo}],
while $\Hadm$ uses the administratively
restricted $\Vadm(d)$ with four channels only.
By construction $\Hgeo(d) \equiv \Hadm(d)$
for all $d < 184$~CE,
and the two series diverge only after the
Yellow Turban~Rebellion onset.

 
$\Hgeo$ incorporates Channel~5
[Eq.~\eqref{eq:C5}],
which after 184~CE makes cross-boundary routes
progressively more expensive
($m_\mathrm{inter}(d) \in [5,\,25]$,
Table~\ref{tab:C5_schedule}),
reflecting the collapse of Han authority over
inter-regional movement.
The effective weight of a geographic edge is:
\begin{equation}
  w_\mathrm{Geo}(e, d, \xi)
  = \wlcp(e)\cdot C_1 \cdot C_2
    \cdot C_3 \cdot C_4 \cdot C_5.
  \label{eq:wGeo}
\end{equation}
The combined effect of Eqs.~\eqref{eq:Vgeo_post}
and~\eqref{eq:wGeo} is:
geographic nodes that physically persisted
after the political dissolution are present
in the network, but their inter-domain
connections are  penalized,
reflecting the  difficulties to connect nodes during troubles times. 

In conclusion, we introduce  geodesic entropy at two spatial scales. At the regional level, 
$H_{\text{geo}}(d)$ is built over all 392 Guo and Houguo 
nodes. At the county level, $H^{\text{xian}}(d)$ is computed 
over all 2,337 Xian nodes  
resolving sub-imperial organization. 
\subsection{\label{sec:Htrd}Trade network:
            \texorpdfstring{$\Htrd$}{H\_Trade}}
 
The trade network encodes the
\emph{exterior commercial connectivity}
of the Han empire. It consists in 
 81~waypoints and 86~segments. 
 The Silk Road node set is time-invariant:
\begin{equation}
  \Vtrd(d) = V_\mathrm{SR},
  \qquad |V_\mathrm{SR}| = 81,
  \quad \forall\, d \in [-200,\, 220]~\mathrm{CE}.
  \label{eq:Vtrd}
\end{equation}
The 86 road edges weights are modulated by
Channels~1--4 but not by Channel~5:
\begin{equation}
  w_\mathrm{Trade}(e,d,\xi) =
  \wlcp(e)\cdot C_1 \cdot C_2 \cdot C_3 \cdot C_4.
  \label{eq:wTrade}
\end{equation}
 

 

\subsection{\label{sec:Hcomb_construction}%
  The combined resilience index $\Hcomb(t)$:
  a time-dependent synthesis}


The three Han networks answer complementary questions.
$\Hadm$ measures cycle redundancy of the formal
prefectural hierarchy---the state's capacity to
move resources through administrative channels.
$\Htrd$ measures cycle redundancy of the Silk~Road
exterior trade network---the state's access to
long-distance commercial revenue.
$\Hgeo$ after 184~CE measures geographic
reorganization of the Three~Kingdoms space,
which belongs to the successor polities, not
to the Han state.

Neither $\Hadm$ nor $\Htrd$ alone captures the
full mobilizable resilience of the Han imperial
system: $\Hadm$ ignores Silk~Road commercial revenue,
which became a primary fiscal source after Zhang~Qian
opened the western routes ($-139$~CE, effective $-120$~CE);
$\Htrd$ is structurally fixed (81~nodes, 86~edges)
and cannot by itself reflect the state's changing
capacity to extract value from those routes.

We therefore define a \emph{combined resilience
index} $\Hcomb(t)$ that measures the topological
complexity the Han state can actually
\emph{mobilize} to respond to shocks:
\begin{equation}
  \Hcomb(t) = [1 - w(t)]\,\Hadm(t)
             + w(t)\,\alpha(t)\,\Htrd(t),
  \label{eq:Hcomb_full}
\end{equation}
where $w(t)$ is the \emph{trade-dependency weight}
and $\alpha(t)$ is the \emph{commercial capture
efficiency}.
This formulation is structurally analogous to
the combined resilience index of Paper~II~\cite{paper2}, with $\Htrd$ playing the role
of $H_{\rm eco}$ and $\alpha(t)$ encoding
Han-specific historical shocks.
$w(t)$ is defined as:
\begin{equation}
  w(t) = w_0
  + w_{\rm SR}\,\sigma(t;\,-120,\,15)
  + w_{\rm WM}\,\sigma(t;\,+9,\,5)\,
    [1-\sigma(t;\,+40,\,10)]
  + w_{\rm col}\,\sigma(t;\,+184,\,8),
  \label{eq:w_trd}
\end{equation}
where $\sigma(t;\tau,\delta) =
[1+e^{-(t-\tau)/\delta}]^{-1}$ is a logistic
transition
and three historically grounded transitions are
encoded (Table~\ref{tab:Hcomb_params}):
(i) the \emph{Silk~Road opening} ($\tau=-120$~CE,
$w_{\rm SR}=0.20$): Zhang~Qian's missions
($-139$~CE) opened the western routes, whose
commercial revenue became fully integrated into
Han fiscal capacity by approximately $-120$~CE,
estimated at $\sim\!15$--20\,\% of total fiscal
receipts at its peak~\cite{Yu1967};
(ii) the \emph{Wang~Mang crisis} ($\tau=+9$~CE,
$w_{\rm WM}=0.15$): the Xin~interregnum
interrupted the administrative network
(node count: $145\to60$), making the state
temporarily more dependent on the 
Silk~Road;the weight returns to baseline at the
Eastern~Han restoration ($\tau = +40$~CE);
and (iii) the \emph{terminal fragmentation}
($\tau=+184$~CE, $w_{\rm col}=0.60$): as the
Yellow~Turban~Rebellion dissolved the
administrative network ($75\to19$ nodes,
$+150\to+220$~CE), the state became
 dependent on Silk~Road revenue
as its last mobilizable topological resource,
with $w(+220) = 0.843$.


The Silk~Road (SR) topology will not change in time during all Han period but its capacity to to extract
fiscal and logistical value from it varies with
control of the Western~Regions (\textit{Xiyu}). To modelize it, we define:
\begin{equation}
  \alpha(t) = \alpha_0\,\hat\alpha(t),
  \label{eq:alpha_trd}
\end{equation}
where $\hat\alpha(t)$ encodes
four historical transitions:
\begin{align}
  \hat\alpha(t) &= 0.20
    + 0.80\,\sigma(t;\,-120,\,12) \notag\\
    &\quad - 0.65\,\sigma(t;\,+9,\,5)\,
      [1-\sigma(t;\,+35,\,8)] \notag\\
    &\quad + 0.40\,\sigma(t;\,+73,\,10)\,
      [1-\sigma(t;\,+160,\,15)] \notag\\
    &\quad - 0.55\,\sigma(t;\,+184,\,8).
  \label{eq:alpha_base}
\end{align}
The four terms describes the following events~\cite{Yu1967}:
\begin{itemize}
\item[i]~gradual SR state revenue under Zhang~Qian and Wu~Di
($-120$~CE onset, $\hat\alpha$ from 0.20 to near~1.0);
\item[ii]~Wang~Mang loss of Xiyu ($+9$~CE,
$-0.65$ reduction~\cite{Yu1967});
\item[iii]~Ban~Chao recovery ($+73$~CE,
50~oasis states reconnected by $+94$~CE~\cite{HouHanShu},
$+0.40$ partial restoration);
\item[iv]~Yellow~Turban interior blockage ($+184$~CE,
$-0.55$, severing the interior distribution network
from the western terminus even as the Silk~Road
continued operating~\cite{deCrespigny2010}).
\end{itemize}
The overall scale $\alpha_0$ is determined by
a fixing $ 
  \Hcomb(220~\text{CE}) = 0.5241=\Hstar$
where $\Hstar$ is interpreted (see Ref.~\cite{paper2} as the topological
irreversibility threshold.
Substituting $\Hadm(220)=0$, $w(220)=0.843$,
$\hat\alpha(220)=0.463$, and $\Htrd(220)=1.228$, one gets $ 
  \alpha_0 = 1.092.$

\begin{table}[!tbp]
\footnotesize
\setlength{\tabcolsep}{4pt}
\caption{\label{tab:Hcomb_params}%
  Parameters of $\Hcomb(t)$ and their historical
  sources.
  $\alpha_0$ is the sole calibrated parameter,
  fixed by the calibration $\Hcomb(220)=\Hstar$.
  All other parameters are fixed independently
  from historical sources.}
\begin{ruledtabular}
\begin{tabular}{c|c|c|c}
Parameter & Symbol & Value & Source \\
\hline
Initial capture eff. & $\alpha_0$ & $1.092$ &
  Calibrated: $\Hcomb(220)=\Hstar$ \\
Pre-SR baseline & $\hat\alpha_{\rm pre}$ & $0.20$ &
  Proto-routes pre-Zhang~Qian~\cite{Yu1967} \\
SR opening amp. & & $+0.80$ &
  Zhang~Qian $-139$, full by $-120$~CE~\cite{Yu1967} \\
Wang~Mang shock & & $-0.65$ &
  Xiyu lost $+9$~CE, restored $+35$~CE~\cite{Yu1967} \\
Ban~Chao recovery & & $+0.40$ &
  50 states $+73$--$+94$~CE~\cite{HouHanShu} \\
Terminal shock & & $-0.55$ &
  YT interior blockage $+184$~CE~\cite{deCrespigny2010} \\
SR opening $w$ & $w_{\rm SR}$ & $0.20$ &
  SR $\sim$15--20\% fiscal receipts~\cite{Yu1967} \\
Wang~Mang $w$ & $w_{\rm WM}$ & $0.15$ &
  Admin collapse, SR dependence~\cite{deCresp2007} \\
Terminal $w$ & $w_{\rm col}$ & $0.60$ &
  Admin$\to 0$: trade is last resource~\cite{deCrespigny2010} \\
\end{tabular}
\end{ruledtabular}
\end{table}


The  $\Hcomb$ trajectory shows four
historically interpretable features.
(Table~\ref{tab:Hcomb_series} lists $\Hcomb$
at key decades with bootstrap-propagated 95\%~CI.)
\begin{itemize}

\item Late Western Han zenith.
$\Hcomb$ peaks at $2.470$ at $t = 0$~CE,
the state derives $\sim\!73\%$
of its mobilizable resilience from $\Hadm$
and $\sim\!27\%$ from Silk~Road capture.

\item  Wang~Mang disruptions.
$\Hcomb$ drops to a local minimum of $1.033$
at $t = +30$~CE---the nadir of the Wang~Mang
disruption.
 
 \item Guangwu~Di restoration show a recovery  to $\Hcomb \approx 1.55$ and corresponds to the re-establishment of the administrative
network (node count: $45\to68$ by $+80$~CE)
pulling the system back from near-irreversibility.

\item  Phase~G terminal decline.
$\Hcomb$ declines overall from $+150$~CE
onward, with a small transient increase at $+180$~CE.
The OLS regression over phase~G
($t\in[+146,+220]$~CE):
\begin{equation}
  \hat\beta_1^{\,\Hcomb} = -0.02560~\text{yr}^{-1},
  \quad R^2 = 0.953,
  \quad p = 3.3\times10^{-5}.
  \label{eq:phaseG_Hcomb}
\end{equation}
 $R^2 = 0.953$ reflects a 
linear terminal decline.
This linearity arises because $w(t)$
smoothly transfers weight from the collapsing
$\Hadm$ to the declining but more stable
$\Htrd$.

\item  Calibration at 220~CE.
$\Hcomb(220) = \Hstar = 0.5241$ by construction.
The bootstrap-propagated 95\%~CI at 220~CE
is $[0.473,\,0.728]$.
\end{itemize}
\begin{table}[H]
\footnotesize
\setlength{\tabcolsep}{4pt}
\caption{\label{tab:Hcomb_series}%
  $\Hcomb(t)$ at selected decades with
  bootstrap-propagated 95\%~CI.
  $w(t)$: trade-dependency weight.
  $\alpha(t) = \alpha_0\,\hat\alpha(t)$:
  effective commercial capture efficiency.
  The Wang~Mang near-miss (top block) and
  phase~G terminal decline (bottom block)
  are shown; the full series is in the supplementary data.}
\begin{ruledtabular}
\begin{tabular}{r D{.}{.}{4} l D{.}{.}{3} D{.}{.}{3}}
$t$ (CE)
  & \multicolumn{1}{c}{$\Hcomb$}
  & CI$_{95}$
  & \multicolumn{1}{c}{$w$}
  & \multicolumn{1}{c}{$\alpha$} \\
\hline
$0$    & 2.4697 & $[2.22,\, 2.56]$ & 0.271 & 0.993 \\
$+30$  & 1.0335 & $[0.87,\, 1.21]$ & 0.358 & 0.642 \\
$+50$  & 1.5499 & $[1.26,\, 1.66]$ & 0.290 & 1.038 \\
\hline
$+150$ & 2.2682 & $[1.83,\, 2.32]$ & 0.258 & 1.092 \\
$+160$ & 1.9590 & $[1.63,\, 2.22]$ & 0.278 & 1.092 \\
$+170$ & 1.7257 & $[1.67,\, 2.22]$ & 0.339 & 1.092 \\
$+180$ & 1.8508 & $[1.52,\, 2.01]$ & 0.477 & 0.956 \\
$+190$ & 1.2742 & $[1.15,\, 1.44]$ & 0.658 & 0.736 \\
$+200$ & 0.8776 & $[0.84,\, 1.11]$ & 0.778 & 0.591 \\
$+210$ & 0.7237 & $[0.60,\, 0.89]$ & 0.828 & 0.529 \\
$+220$ & 0.5241 & $[0.47,\, 0.73]$ & 0.843 & 0.506 \\
\end{tabular}
\end{ruledtabular}
\end{table}

%
 
 
\begin{table}[H]
\scriptsize
\caption{\label{tab:network_comparison}%
  Defining properties of the three Han networks. Administrative activation:
  Eq.~\eqref{eq:V_admin}; geographic persistence: Eq.~\eqref{eq:Vgeo_post};
  Channel~5: Sec.~\ref{sec:C5}.}
\resizebox{\textwidth}{!}{%
\begin{tabular}{llll}
\toprule
\multicolumn{4}{l}{\textit{Panel A -- Network structure}} \\
Property & $\Hadm$ & $\Hgeo$ & $\Htrd$ \\
\midrule
Node catalog & 392 Guo + Houguo & 392 Guo + Houguo & 81 Silk Road waypoints \\
Activation & CHGIS active window & same as $\Hadm$ before 184; extended to 220 after 184 & constant \\
$N_{\min}$ / $N_{\max}$ & 19 / 145 & 19 / 392 & 81 / 81 \\
Edge catalog & Interior LCP (54,522) & Interior LCP (same catalog) & Silk Road (86) \\
Friction channels & $C_1,C_2,C_3,C_4$ & $C_1,C_2,C_3,C_4,C_5$ & $C_1,C_2,C_3,C_4$ \\
\midrule
\multicolumn{4}{l}{\textit{Panel B -- Temporal behavior and signal}} \\
Divergence & reference series & 184~CE (Yellow Turban Rebellion) & independent from inception \\
Meaning & formal Han authority & territory and post-184 reorganization & commercial connectivity \\
Phase~G slope & $-0.0261$~yr$^{-1}$ ($R^2=0.689$, $p=0.0108$) & $+0.0169$~yr$^{-1}$ ($R^2=0.636$, $p=0.018$) & $-0.0020$~yr$^{-1}$, n.s. \\
Value at 220~CE & $\Hadm=0.000$ & $\Hgeo=3.065$ & $\Htrd=1.228$ \\
\bottomrule
\end{tabular}}
\end{table}
 
Table~\ref{tab:hfull_bias} summarizes why the preliminary full-network observable is not used as a primary result.

\begin{table}[!tbp]
\footnotesize
\setlength{\tabcolsep}{4pt}
\caption{\label{tab:hfull_bias}%
  Quantification of the Silk Road floor effect
  in the combined ``full'' network
  (preliminary analysis, not used in the present paper).
  $\phi_\mathrm{SR}$: fraction of active full-network
  nodes contributed by the Silk Road
  defined as the active Silk Road node fraction.
  As $\phi_\mathrm{SR} \to 1$ during the terminal collapse,
  $\Hfull \to \Htrd$,
  masking the administrative collapse signal.}
\begin{ruledtabular}
\begin{tabular}{r D{.}{.}{3} D{.}{.}{3} D{.}{.}{3} D{.}{.}{3}}
$d$ (CE)
  & \multicolumn{1}{c}{$\phi_\mathrm{SR}$}
  & \multicolumn{1}{c}{$\Hfull$}
  & \multicolumn{1}{c}{$\Hadm$}
  & \multicolumn{1}{c}{$\Htrd$} \\
\hline
$-10$ & 0.358 & 2.667 & 2.724 & 1.119 \\
$ 20$ & 0.574 & 1.534 & 2.074 & 1.527 \\
$100$ & 0.519 & 1.458 & 1.979 & 1.171 \\
$150$ & 0.519 & 1.523 & 2.565 & 1.296 \\
$180$ & 0.536 & 1.306 & 2.188 & 1.548 \\
$210$ & 0.681 & 1.449 & 1.034 & 1.246 \\
$220$ & 0.810 & 1.300 & 0.000 & 1.228 \\
\end{tabular}
\end{ruledtabular}
\end{table}
 
 \begin{figure}[!tbp]
     \centering
     \includegraphics[width=0.82\linewidth]{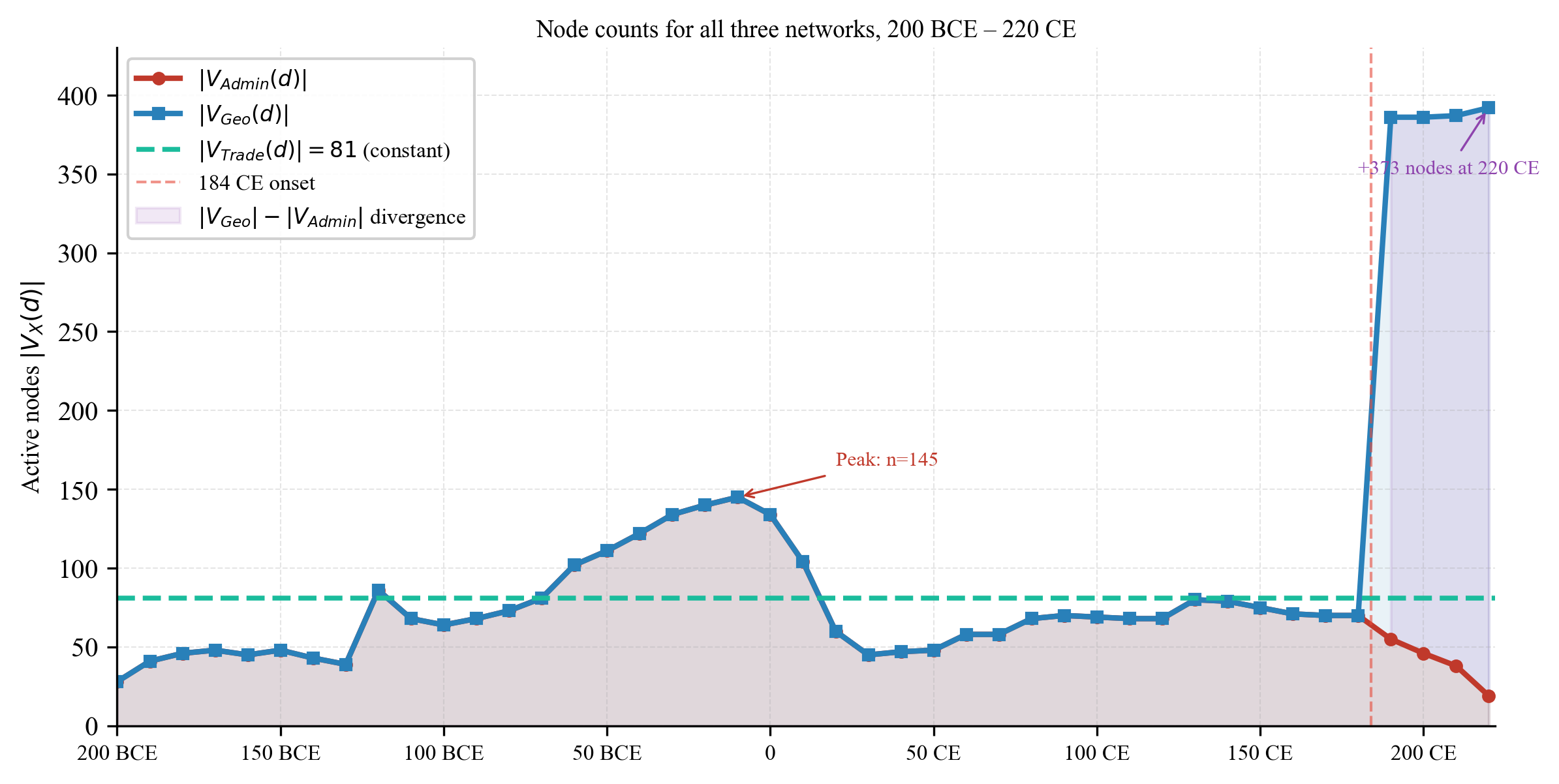}
     \caption{\label{fig:node_counts}%
    Active node counts for the three networks,
    200~BCE to 220~CE.
    Administrative and geographic networks are
    identical until $d = 184$~CE (Yellow Turban
    Rebellion), after which the geographic
    network retains all historically active
    Guo prefectures while the administrative
    network continues to contract.}
 \end{figure}

 \begin{figure}[!tbp]
     \centering
     \includegraphics[width=0.82\linewidth]{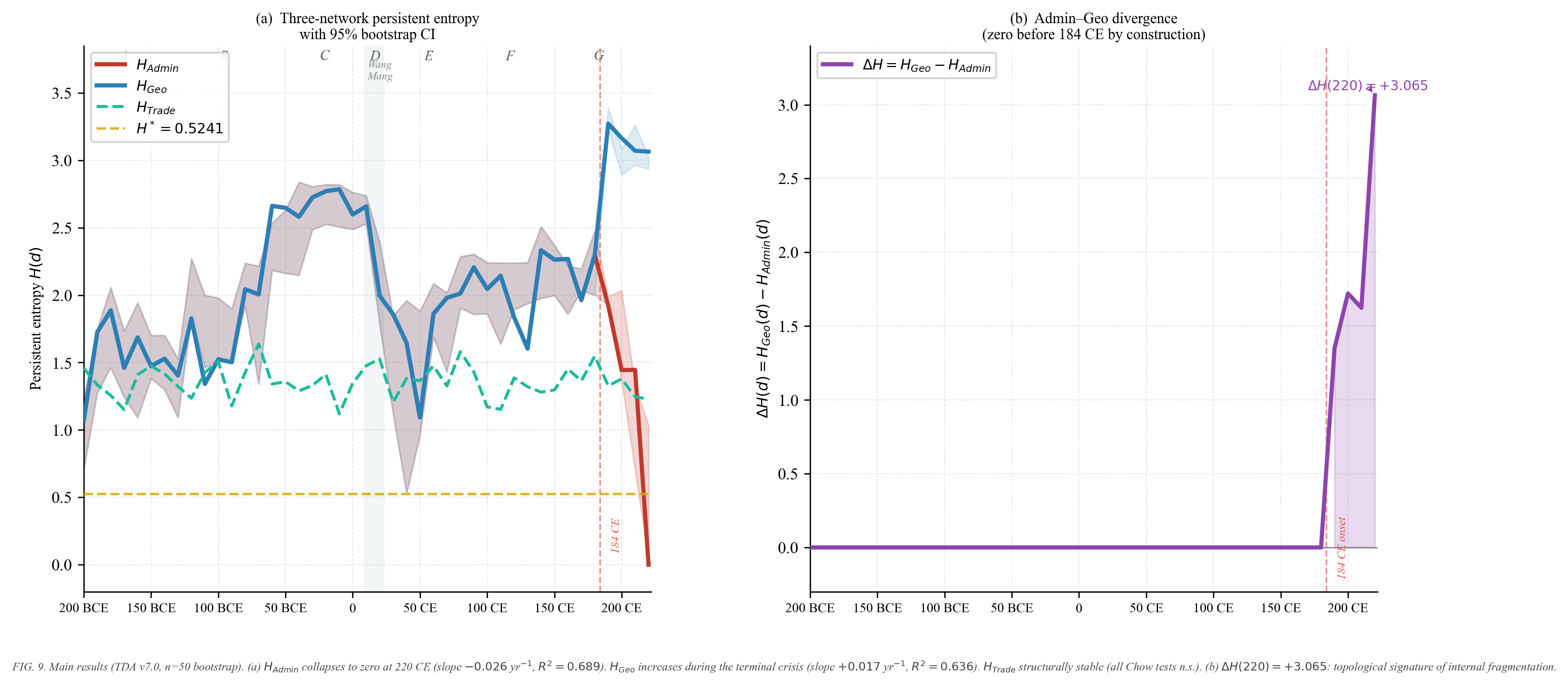}
     \caption{\label{fig:three_networks}%
    The three-network decomposition.
    (a)~Persistent entropy time series
    for all three networks with 95\% bootstrap CI.
    (b)~Admin--Geo divergence
    $\DHH(d) = \Hgeo(d) - \Hadm(d)$,
    identically zero before 184~CE and
    rising to $+3.065$ at the formal
    Han dissolution in 220~CE.}
 \end{figure}


\section{\label{sec:results}Results}


\subsection{\label{sec:res_admin}Administrative network:
            \texorpdfstring{$\Hadm$}{H\_Admin}}


Figure~\ref{fig:three_networks} shows $\Hadm(d)$
over the full 420-year period.
The series rises from $\Hadm(-200) = 1.076$
(28 active prefectures, two connected components in the early Western Han)
to a peak of $\Hadm(-10) = 2.785$
(145~prefectures, 27 finite-lifetime B$_1$ cycles,
filtration threshold $\varepsilon_{-10} = 619$~km-equivalents),
followed by a sharp and sustained decline
terminating at $\Hadm(220) \approx 0$
(19~prefectures, one residual B$_1$ cycle).
The full range $H \in [0.000, 2.785]$ is
illustrating the well-known evolution of the
Han administrative network from a fragmented
early state to a topologically integrated empire
and back to minimal connectivity.

\subsubsection{Phase regressions}

Table~\ref{tab:phase_regressions} reports
the OLS phase regressions for all seven
Han historical phases.
Two phases show statistically significant trends:

\emph{Phase~B (Wu~Di expansion,
$-141$ to $-50$~BCE):}
$\hat\beta_1 = +0.0132$~yr$^{-1}$,
$R^2 = 0.674$, $p = 0.004$.
This is the only phase of statistically
significant topological \emph{growth} in
$\Hadm$, corresponding to the maximum expansion of the Han administration (opening
of the Hexi Corridor, the Zhang~Qian missions,
the "Silk Road" Protectorate).
The strong fit ($R^2 = 0.674$) indicates that
this expansion was a sustained, approximately
linear process rather than a step change.

\emph{Phase~G (Late Han crisis,
$+146$ to the end:}
$\hat\beta_1 = -0.0261$~yr$^{-1}$,
$R^2 = 0.689$, $p = 0.0108$.
This is the collapse phase.
The magnitude of the slope is
$13.4\times$ larger than the corresponding
collapse slope reported in Paper~I~\cite{paper1}
($-2.30 \times 10^{-3}$~yr$^{-1}$),
confirming that the Han administrative collapse
is substantially more rapid than that of comparable historical systems.
The statistical fit is comparable
($R^2 = 0.689$),
indicating that both collapses follow
approximately linear trends in $H$ during
their terminal phases, but at very different rates.

The remaining phases (A, C, E, F) show no
statistically significant trends,
consistent with a system with
periods of administrative stasis and adjustment
interrupted by the two major expansion
(phase~B) and collapse (phase~G) events.
Phase~D (Wang Mang interregnum, 9--23~CE)
has $n = 2$~decades, insufficient for OLS
regression, but its topological signature
appears clearly in the Chow test results below.

\subsubsection{Chow structural break tests}

Table~\ref{tab:chow_all} reports the Chow
test  ($F(t^*)$ and $p$-value)
for all eight historical breakpoints.
$\Hadm$ registers statistically significant
structural breaks at seven of the eight
predetermined breakpoints (7/8,
significance at $\alpha = 0.05$):
at $t^* \in \{-33, +9, +25, +89, +166,
              +184, +220\}$~CE.

The only non-significant breakpoint is
$t^* = -141$~CE ($F = 2.11$, $p = 0.135$) which corresponds to 
the transition from early Western Han (phase~A)
to the Wu~Di expansion (phase~B). This is due to the fact that this transision is so  gradual that no structural
break is detectable in the $H(d)$ series,
consistent with the progressive  Han expansion.




\subsection{\label{sec:res_geo}Geographic network:
            \texorpdfstring{$\Hgeo$}{H\_Geo}}


As the unified Han state dissolves,
the geographic space it occupied reorganizes
into locally coherent sub-networks.
Each proto-kingdom  develops
its own internal administrative, trade and
military infrastructure,
generating regional cycles in the
Rips complex that were absent under
the more centralized Han system.
The total topological complexity of the
geographic territory \emph{increases}
because fragmentation generates more
independent loops at the sub-imperial level.



\subsection{\label{sec:res_trade}Trade network:
            \texorpdfstring{$\Htrd$}{H\_Trade}}


$\Htrd(d)$ ranges between $1.119$ and $1.637$
over the full 43-decade analysis period,
with a mean of $\bar{H}_\mathrm{Trade} = 1.357$
and a standard deviation of $0.119$.
No systematic trend is visible in any
of the six historical phases;
all six OLS slopes are statistically
non-significant ($p > 0.08$).
The variation of $H$ 
is  explained by the stochastic
variability of Channel~2 and the
time-varying friction of Channels~1 and~4
acting on a topologically fixed node and edge set.

the Han Silk Road operated at a temporal
and economic scale that transcended the
political cycles of the Han state.
The trade routes were maintained by
the accumulated interests of thousands
of merchants, oasis cities, and steppe
polities.

\subsection{\label{sec:res_topdyn}Topological dynamics:
            Wasserstein velocity and early-warning
            observables}

The persistent entropy $H(d)$ is a scalar
summary of the $\beta_1$ barcode at each
decade.
A complementary perspective is provided by
\emph{how fast} the barcode changes in time
and by \emph{summary indicators} that combine
multiple aspects of the barcode geometry into
a single early-warning score.
This section quantifies the topological
dynamics using three independent indicators:
the Wasserstein velocity
$\dot{W}_2$ (the rate of change of the
persistence diagram),
the cross-network Wasserstein distance
$W_\times(\mathrm{Admin}, \mathrm{Geo})$
(the metric-space distance between
the two barcodes),
and the Integrated Change Tracker (ICT),
a normalized composite of multiple
persistence observables.
All three indicators are independent of
the entropy $H(d)$ and provide consistent,
mutually reinforcing evidence for an
early-warning signal in the Admin network
beginning at $d = 170$~CE.

\subsubsection{\label{sec:res_Wvelocity}%
   Rate of topological change}

The $L^2$ Wasserstein distance $W_2(d,\,d{+}10)$
measures how much the set of topological cycles
changed between two consecutive decades \cite{Edelsbrunner2002}.
A large $W_2$ signals a substantial reorganization
of the cycle structure; a small $W_2$ signals
near-stability.
The per-year Wasserstein velocity is:
\begin{equation}
  \dot{W}_2(d) = \frac{W_2(d,\, d{+}10)}{10}
  \qquad [\mathrm{filtration~units~yr}^{-1}],
  \label{eq:Wvelocity}
\end{equation}
estimated at the midpoint
$t_\mathrm{mid} = d + 5$.
$\dot{W}_2$ measures the
\emph{topological acceleration} of the network.

\paragraph{Baseline.}
The smoothed velocity $\dot{W}_2^\mathrm{sm}(t)$
(3-point running mean) stays near a stable baseline
of $\approx 3.07$~filtration units~yr$^{-1}$
throughout the unified Han period ($t < 100$~CE),
before rising sharply as the collapse approaches.

\paragraph{Early-warning signal in the Admin network.(see Table~\ref{tab:Wvelocity})}
The peak value, $\dot{W}_2 = 192$~yr$^{-1}$ is obtained 
at $t = 215$~CE and  is $63\times$ the baseline
and represents the single largest
decade-to-decade topological change
in the 420-year record, 5 years before the formal dissolution.


\begin{table}[!tbp]
\footnotesize
\setlength{\tabcolsep}{4pt}
\caption{\label{tab:Wvelocity}%
  Wasserstein velocity $\dot{W}_2(t)$
  for the Admin network at key decades,
  as a multiple of the pre-crisis baseline
  $\dot{W}_2^\mathrm{base} = 3.07$~yr$^{-1}$.
  Values are smoothed (3-point running mean).
  The early-warning signal ($>2\times$ baseline)
  appears at $t = 175$~CE, 45~years before
  formal dissolution.}
\begin{ruledtabular}
\begin{tabular}{r r r l}
$t$ (CE) & $\dot{W}_2$ & $\dot{W}_2/\dot{W}_2^\mathrm{base}$
  & signal level \\
\hline
$-195$ &   6.2 &  2.0$\times$ & fluctuation \\
$175$  &  13.4 &  4.3$\times$ & \textbf{early warning} \\
$185$  &  17.1 &  5.6$\times$ & elevated \\
$195$  &  43.9 & 14.3$\times$ & acute \\
$205$  & 106.7 & 34.7$\times$ & critical \\
$215$  & 192.3 & 62.6$\times$ & \textbf{terminal peak} \\
\end{tabular}
\end{ruledtabular}
\end{table}

\subsubsection{\label{sec:res_Wcross}%
  Cross-network Wasserstein distance}

The cross-network Wasserstein distance
\begin{equation}
  W_\times(d) = W_2\!\bigl(
    \mathcal{D}_d^\mathrm{Admin},\;
    \mathcal{D}_d^\mathrm{Geo}
  \bigr)
  \label{eq:Wcross}
\end{equation}
compares the Admin and Geo persistence diagrams
at the same decade $d$.
 $W_\times(d)$ measures the
full geometric distance between the two barcodes:
it is zero when the two networks are topologically
identical and grows as they differ.

$W_\times(d) = 0$ for all
$d \leq 180$~CE as expected by construction.
At $d = 190$~CE, $W_\times$ jumps
discontinuously to $737$ in a single decade 
confirming that the Admin--Geo separation
is an abrupt topological event rather than
a gradual drift.
$W_\times$ then grows to $987$ (200~CE),
$1{,}085$ (210~CE), and $7{,}948$ at
the end of Han period.

\subsubsection{\label{sec:res_xi}%
  Correlation length divergence:
  topological scale of the Three~Kingdoms}

The \emph{topological correlation length}
$\xi(d)$ is defined as the weighted spread
of bar birth values in the $\beta_1$ diagram:
\begin{equation}
  \xi(d) = \sqrt{\mathrm{Var}\!\left[
    \{b_i\}_{i=1}^{n_d}
  \right]},
  \label{eq:xi_def}
\end{equation}
where $b_i$ are the birth parameters of
the $n_d$ bars at decade $d$.
A large $\xi$ means loops appear at many 
different spatial scales during the same decade --- the network 
has structure at both local and long-range levels. 
As the network transitions from a centralized, 
single-scale organization to a fragmented multi-scale 
one, $\xi$ is expected to grow sharply.

\begin{itemize}
    
\item \textbf{Pre-crisis baseline.}
For $d < 184$~CE, both Admin and Geo share
$\xi_\mathrm{base} \approx 128$ filtration
units averaged over the stable Western Han period (before 100~CE).
This is the characteristic loop scale
of the unified Han network.

\item \textbf{Admin correlation length.}
$\xi_\mathrm{Admin}$ first crosses
$5\times$ baseline at $d = 190$~CE
($\xi = 657$, $5.1\times$),
peaks at $d = 200$~CE
($\xi = 1{,}568$, $12.2\times$),
then declines to $739$ at 210~CE
and to $\xi = 0$ at end of Han period.
$\xi_\mathrm{Admin}$ rises as the
barcode scatters into a few widely-spaced
bars (190--200~CE), then collapses
as those bars are lost one by one.

\item \textbf{Geo correlation length:
           monotonic divergence.}
$\xi_\mathrm{Geo}$ diverges
\emph{monotonically} from 190 to the end of Han period,
reaching $\xi_\mathrm{Geo}(220) = 7{,}105$:
\begin{equation}
  \frac{\xi_\mathrm{Geo}(220)}%
       {\xi_\mathrm{base}}
  = \frac{7{,}105}{128} \approx 55.5.
  \label{eq:xi_ratio}
\end{equation}
%

The divergence of $\xi_\mathrm{Geo}$
is consistent with a topological analogue
of a correlation length divergence
near a phase transition~\cite{Bollobas2001},
in which the geographic network approaches
a new fixed point 
rather than a disordered state.
\end{itemize}

\subsubsection{\label{sec:res_ICT}%
  Integrated Change Tracker:
  early-warning chronology}

The Integrated Change Tracker (ICT) averages
three normalized early-warning indicators.
The \emph{susceptibility} $\chi(d) = \mathrm{Var}(\{\ell_i(d)\})$;
the \emph{correlation length} $\xi(d)$ is the weighted spread of bar birth values, a proxy for the spatial scale of active routing cycles;
and $\dot{W}_2(d)$ is the smoothed Wasserstein velocity
defined in Eq.~\eqref{eq:Wvelocity}.
Their normalized average is:
\begin{equation}
  \mathrm{ICT}(d) =
  \frac{1}{3}\!\left[
    \underbrace{\frac{\chi(d)}{\chi_\mathrm{max}}}_{\text{susceptibility}}
    +
    \underbrace{\frac{\xi(d)}{\xi_\mathrm{max}}}_{\text{correlation length}}
    +
    \underbrace{\frac{\dot{W}_2(d)}{\dot{W}_{2,\mathrm{max}}}}_{\text{velocity}}
  \right],
  \label{eq:ICT_def}
\end{equation}
where each term is normalized to its series maximum.
ICT ranges from 0 (all indicators at baseline)
to 1 (all three simultaneously at their peak),
signalling proximity to a topological transition.

\begin{itemize}
\item \textbf{Admin ICT chronology.}
(see Table~\ref{tab:ICT}) 
Three quantitative thresholds are
identifiable:

\begin{enumerate}
  \item \emph{Early warning} ($\mathrm{ICT} > 0.1$):
    first crossed at $d = 170$~CE,
    $\mathrm{ICT}(170) = 0.118$.
    This is \textbf{50 years before the formal
    dissolution} (220~CE) and \textbf{14 years
    before the Yellow Turban Rebellion} (184~CE).
    At this point, $\Hadm$ shows no
    statistically significant phase~G trend
    ($\hat\beta_1 = -0.0040$~yr$^{-1}$,
    $p = 0.61$ for $d \leq 170$~CE alone).

  \item \emph{Elevated risk} ($\mathrm{ICT} > 0.2$):
    first crossed at $d = 190$~CE,
    $\mathrm{ICT}(190) = 0.280$.
    This coincides with the decade
    after Yellow Turban activation,
    and with the onset of the Admin--Geo
    divergence ($W_\times = 737$,
    $\DHH = +1.35$).
    All four indicators
    ($\Hadm$, $\Hgeo$, $W_\times$, ICT)
    fire simultaneously at this threshold.

  \item \emph{Active collapse} ($\mathrm{ICT} > 0.35$):
    first crossed at $d = 200$~CE,
    $\mathrm{ICT}(200) = 0.402$.
    This decade sees $\Hadm$ lose four
    of its five remaining bars,
    $\xi_\mathrm{Admin}$ reaching its peak,
    and the Binder cumulant $U_4$ crossing
    from negative (disordered) to positive
    (concentrated), consistent with the
    condensation funnel entering its final stage.
\end{enumerate}

\item \textbf{Geo and Trade ICT.}
$\mathrm{ICT}_\mathrm{Geo}$ rises
monotonically from 0.32 (190~CE)
to 0.97 (220~CE): the Geo network
approaches its maximum structural change
at dissolution,
with $\mathrm{ICT}_\mathrm{Geo} \to 1$. The monotonic rise from 0.32 to 0.97 between 190 and 220 CE shows us that  the disintegration of the Han network was not sudden but an accelerating process.
\end{itemize}
\begin{table}[!tbp]
\footnotesize
\setlength{\tabcolsep}{4pt}
\caption{\label{tab:ICT}%
  Integrated Change Tracker (ICT) for the
  Admin network across the terminal phase,
  with the three early-warning thresholds.
  ICT is computed from Eq.~\eqref{eq:ICT_def}
  using susceptibility $\chi$,
  correlation length $\xi$,
  and smoothed Wasserstein velocity
  $\dot{W}_2^\mathrm{sm}$,
  each normalized to their series maximum.
  At $d = 170$~CE, the entropy series $\Hadm$
  shows no significant phase~G trend,
  establishing ICT as an \emph{earlier} indicator.}
\begin{ruledtabular}
\begin{tabular}{r D{.}{.}{4} l}
$d$ (CE) & \multicolumn{1}{c}{ICT} & signal level \\
\hline
$150$ & 0.0270 & background \\
$160$ & 0.0538 & background \\
$170$ & 0.1178 & \textbf{early warning} ($>0.1$) \\
$180$ & 0.1067 & early warning \\
$190$ & 0.2803 & \textbf{elevated risk} ($>0.2$) \\
$200$ & 0.4018 & \textbf{active collapse} ($>0.35$) \\
$210$ & 0.3839 & active collapse \\
\end{tabular}
\end{ruledtabular}
\end{table}

\subsection{\label{sec:res_divergence}Admin--Geo divergence:
            topological signature of internal fragmentation}

$\DHH(d) = \Hgeo(d) - \Hadm(d)$ increases overall during the terminal phase:
$\DHH(190) = +1.349$,
$\DHH(200) = +1.721$,
$\DHH(210) = +1.625$,
$\DHH(220) = +3.065$.
The final value $\DHH(220) = +3.065$ is
the quantitative signature of the
\emph{completion} of the Han collapse:
at the end, the geographic territorial space
has 3.065~entropy units \emph{more}
topological complexity than the
administrative network.
The state is gone ($\Hadm \approx 0$);
the territory is not.

\subsubsection{Relation to B$_0$ fragmentation}
 
A naive expectation would be that
the Three~Kingdoms partition is captured
by $\beta_0 = 3$~components in $\Hgeo$
after $d = 208$~CE.
In practice, $\beta_0 = 1$ throughout
all 43~decades in all three networks:
the Guo-level geographic network
(approximately 170--392~nodes,
with dense LCP-based connections within each region)
is still well connected even when
Channel~5 multiplies inter-domain edges
by $25\times$.
This is historically accurate:
Wei, Shu, and Wu were not hermetically sealed.

 

--
\subsubsection{Three Kingdoms as emergent \texorpdfstring{$\beta_1$}{B1}
               cycles under the fragmentation model}

The Wei, Shu, and Wu cycles emerge
purely from the topological structure of the
post-184~CE geographic network, and correspond to the
three geographically coherent sub-regions of China
that share a natural internal connectivity
much stronger than their cross-boundary connectivity:
(i) the \textit{Wei domain}
(Guanzhong basin, Hebei, Shandong---north and northeast);
(ii) the \textit{Shu domain}
(Sichuan Basin, enclosed by the Qinling and Daba ranges---southwest);
and (iii) the \textit{Wu domain}
(Jiangnan, the Yangzi Delta and eastern coast---southeast).

\paragraph{Why the cycles appear where they do.}
The Qinling--Daba mountain range separating
the Guanzhong/North China Plain from the Sichuan Basin,
the Nanling range separating Jiangnan from the south,
and the Taihang range separating the Wei and Hebei plains.
In the unified Han period, these barriers are
present but not decisive.
After 184~CE,their costs are so high that they represents strict frontiers isolating geographical region. A a result, one observes:

\begin{itemize}
  \item \emph{Three long-lived bars} born at low $\epsilon$
    and persisting to large radii, corresponding to the
    three intra-domain circuits:
    (i) the Guanzhong--Hebei--Shandong loop (Wei),
    (ii) the Sichuan Basin internal loop (Shu), and
    (iii) the Jiangnan--Yangzi loop (Wu).
    These bars carry the majority of the total persistence
    weight and drive the high $\Hgeo(220) = 3.065$.

  \item \emph{A larger population of short-lived bars}
    born and dying at intermediate $\epsilon$,
    corresponding to sub-regional circuits within each
    domain and to the few persistent inter-domain
    connections (e.g., the Yangzi corridor linking Wu and Shu
    through the Gorges, and the Wei--Han valley crossing
    between Wei and Shu domains).
\end{itemize}




\paragraph{Quantitative support.}
The transition is visible in the entropy series:
$\Hgeo$ rises from $2.299$ at $d = 180$~CE
(single-regime unified structure) to $3.272$ at
$d = 190$~CE in a single decade,
the largest single-decade increase in the entire
420-year series.
The bootstrap CI for $\Hgeo(190)$ is $[3.232, 3.390]$,
entirely above the pre-184 range,
confirming the abruptness of the topological reorganization.
This coincides with the Yellow~Turban~Rebellion
onset\cite{deCrespigny2010}. 

\subsubsection{Relationship to $\beta_0$ atomization.}

At the Xian (county) level, the picture is different:
$\beta_0^{\rm Xian}$ explodes from 294 to 1,174
at $d = 190$~CE (Table~\ref{tab:b0_xian}),
revealing that at the county scale,
the Han empire first \emph{atomized}---
fragmented below the prefecture scale into
more than one thousand isolated county clusters

The $\beta_0$ atomization and the $\beta_1$ formation of three structure seems at first sight as contradictory. But in fact, these two pictures are complementary. 
At the county level, $H_{\text{xian}}$ captures the dissolution 
of the fine-grained administrative fabric: the breakdown of local connections 
 and counties isolation. At the regional level, 
the $\beta_1$ cycles capture a simultaneous reorganization where we can see the 
the surviving long-range routes consolidate into three coherent 
geographic domains. 

\subsubsection{Quantification of the topological
               gap: signal-to-noise ratio of the
               Three~Kingdoms cycles}

The identification of the Three~Kingdoms as
emergent $\beta_1$ cycles rests not only on their
existence but on their \emph{quantitative
separation} from the background of short-lived
bars in the persistence diagram.
We measure this separation via the
\emph{top-3 gap ratio}:
\begin{equation}
  G_3(d) =
  \frac{\bar\delta_{[1,3]}(d)}{\bar\delta_{[4,n]}(d)},
  \label{eq:gap_ratio}
\end{equation}
where $\bar\delta_{[1,3]}$ is the mean lifetime
of the three longest-lived bars and
$\bar\delta_{[4,n]}$ is the mean lifetime of
all remaining bars at decade $d$.
A large $G_3$ indicates that the three dominant
cycles are clearly distinguished from the
``noise'' of sub-regional and local circuits.

\paragraph{The gap appears abruptly at 184 CE.}
At $d = 170$~CE (the last pre-Yellow Turban decade),
$G_3 = 4.9\times$: the top-3 bars are about
five times more persistent than the rest,
a typical background level for a unified network.
At $d = 180$~CE: $G_3 = 3.2\times$
(the diagram is relatively homogeneous---
13~bars of similar lifetime).
At $d = 190$~CE---the first post-184~CE decade---
$G_3$ jumps to $\mathbf{10.9\times}$.
This is the moment when the Three~Kingdoms
cycles become topologically unambiguous.
The jump from 3.2 to 10.9 in a single decade
is the sharpest change in the gap ratio across
the entire 420-year series.

\paragraph{Stability of the gap 190--220 CE.}
The gap ratio then stabilizes at approximately
$G_3 \approx 12\times$ across the entire
190--220~CE period:
$G_3 = 10.9$ (190~CE), $12.9$ (200~CE),
$11.8$ (210~CE), $12.4$ (220~CE).
This stability is the key evidence that the
three dominant cycles are  a  structural feature
of the geographic network throughout the
Three~Kingdoms incubation period.



\begin{table}[H]
\footnotesize
\setlength{\tabcolsep}{4pt}
\caption{\label{tab:geo_gap}%
  Top-3 gap ratio $G_3(d)$ and weight
  concentration in the Geo $\beta_1$ barcode
  at terminal decades.
  $n$: total bars. $G_3$: mean lifetime of top-3
  bars divided by mean lifetime of remaining bars
  [Eq.~\eqref{eq:gap_ratio}].
  $w_3$: fraction of total persistence weight
  carried by the top-3 bars.
  $\delta_1$: lifetime of the single most persistent
  bar (the dominant intra-domain circuit).
  At $d = 180$~CE (pre-divergence baseline):
  $n = 13$, $G_3 = 3.2\times$, no Three~Kingdoms
  signal present.}
\begin{ruledtabular}
\begin{tabular}{r r r r r}
$d$ (CE) & $n$ & $G_3$ & $w_3$ & $\delta_1$ \\
\hline
$180$ & 13 &  3.2$\times$ & 0.492 &      166 \\
$190$ & 61 & 10.9$\times$ & 0.361 &    5{,}953 \\
$200$ & 61 & 12.9$\times$ & 0.400 &    8{,}161 \\
$210$ & 53 & 11.8$\times$ & 0.414 &    7{,}134 \\
$220$ & 54 & 12.4$\times$ & 0.421 &   27{,}160 \\
\end{tabular}
\end{ruledtabular}
\end{table}


The phase-regression table below collects the slope estimates used in the preceding discussion of Admin, Geo, and Trade behavior.

\begin{table*}[t]
\scriptsize
\setlength{\tabcolsep}{3pt}
\caption{\label{tab:phase_regressions}%
  OLS phase regression results for $\Hadm$,
  $\Htrd$, and $\Hgeo$.
  Phases~A through~F are identical for
  $\Hadm$ and $\Hgeo$ (by construction,
  since $\Hgeo \equiv \Hadm$ before 184~CE);
  only their phase~G slopes differ.
  Phase~D is excluded from OLS ($n=2$; see
  Table~\ref{tab:phases}).
  $\hat\beta_1$: slope in units of
  entropy per year.
  $R^2$: coefficient of determination.
  Significance: $^*p<0.05$, $^{**}p<0.01$,
  $^{***}p<0.001$; (n.s.) otherwise.
  For phases A--F, values in the Geo column
  are identical to Admin (marked $\equiv$).}
\begin{ruledtabular}
\begin{tabular}{l l
    D{.}{.}{6} D{.}{.}{3} r l
    D{.}{.}{6} D{.}{.}{3} r l
    D{.}{.}{6} D{.}{.}{3} r l }
& &
\multicolumn{4}{c}{$\Hadm$} &
\multicolumn{4}{c}{$\Htrd$} &
\multicolumn{4}{c}{$\Hgeo$} \\
\cline{3-6}\cline{7-10}\cline{11-14}
Ph. & Period (CE)
  & \multicolumn{1}{c}{$\hat\beta_1$}
  & \multicolumn{1}{c}{$R^2$}
  & \multicolumn{1}{c}{$n$}
  & sig.
  & \multicolumn{1}{c}{$\hat\beta_1$}
  & \multicolumn{1}{c}{$R^2$}
  & \multicolumn{1}{c}{$n$}
  & sig.
  & \multicolumn{1}{c}{$\hat\beta_1$}
  & \multicolumn{1}{c}{$R^2$}
  & \multicolumn{1}{c}{$n$}
  & sig. \\
\hline
A & $-206$ to $-141$
  & 0.00411 & 0.074 & 6 & (n.s.)
  & 0.00058 & 0.007 & 6 & (n.s.)
  & -  & - &- & \\
B & $-141$ to $-50$
  & 0.01324 & 0.674 & 10 & $^{**}$
  & 0.00078 & 0.032 & 10 & (n.s.)
  & -  & - &- & \\
C & $-50$  to $+8$
  & 0.00117 & 0.062 & 6 & (n.s.)
  & -0.00139& 0.066 & 6 & (n.s.)
  &-  &- &- & \\
E & $+25$  to $+88$
  & 0.00725 & 0.156 & 6 & (n.s.)
  & 0.00519 & 0.575 & 6 & (n.s.)
  & -  & - & - & \\
F & $+89$  to $+145$
  & -0.00285& 0.040 & 6 & (n.s.)
  & -0.00021& 0.001 & 6 & (n.s.)
  & -  & - &- & \\
\hline
G & $+146$ to $+220$
  & -0.02606& 0.689 & 8 & $^{**}$
  & -0.00200& 0.212 & 8 & (n.s.)
  & +0.01690& 0.636 & 8 & $^{*}$  \\
\end{tabular}
\end{ruledtabular}
\end{table*}

The corresponding Chow tests are reported immediately after the phase regressions to keep the structural-break evidence attached to the text.

\begin{table*}[t]
\scriptsize
\setlength{\tabcolsep}{3pt}
\caption{\label{tab:chow_all}%
  Chow structural break test results
  for all three networks at the eight
  \emph{a priori} Han historical breakpoints.
  $F$: Chow statistic.
  Significance: $^*p<0.05$, $^{**}p<0.01$,
  $^{***}p<0.001$; (n.s.) not significant.
  Summary row: number of significant breaks.
  Note that the breakpoint $-141$~CE is not
  significant for any network, while $+9$~CE
  and $+25$~CE (Wang~Mang) are significant
  for both $\Hadm$ and $\Hgeo$ but not
  for $\Htrd$.}
\begin{ruledtabular}
\begin{tabular}{r
    D{.}{.}{2} l D{.}{.}{6} l
    D{.}{.}{2} l D{.}{.}{6} l
    D{.}{.}{2} l D{.}{.}{6} l}
& \multicolumn{4}{c}{$\Hadm$}
& \multicolumn{4}{c}{$\Htrd$}
& \multicolumn{4}{c}{$\Hgeo$} \\
\cline{2-5}\cline{6-9}\cline{10-13}
$t^*$ (CE)
  & \multicolumn{1}{c}{$F$} & sig.
  & \multicolumn{1}{c}{$p$} &
  & \multicolumn{1}{c}{$F$} & sig.
  & \multicolumn{1}{c}{$p$} &
  & \multicolumn{1}{c}{$F$} & sig.
  & \multicolumn{1}{c}{$p$} & \\
\hline
$-141$ & 2.11 & (n.s.) & 0.135 &
        & 0.17 & (n.s.) & 0.843 &
        & 0.17 & (n.s.) & 0.845 & \\
$ -33$ & 12.35 & $^{***}$& 0.000070 &
        & 0.19 & (n.s.) & 0.831 &
        & 1.88 & (n.s.) & 0.166 & \\
$  +9$ & 13.91 & $^{***}$& 0.000028 &
        & 0.62 & (n.s.) & 0.541 &
        & 12.88& $^{***}$& 0.000051 & \\
$ +25$ & 12.51 & $^{***}$& 0.000063 &
        & 0.27 & (n.s.) & 0.767 &
        & 19.07& $^{***}$& 0.0000020 & \\
$ +89$ & 5.89  & $^{**}$ & 0.005802 &
        & 0.95 & (n.s.) & 0.396 &
        & 3.01 & (n.s.) & 0.061 & \\
$+166$ & 10.82 & $^{***}$& 0.000183 &
        & 1.12 & (n.s.) & 0.337 &
        & 3.62 & $^{*}$  & 0.036 & \\
$+184$ & 12.45 & $^{***}$& 0.000066 &
        & 0.83 & (n.s.) & 0.444 &
        & 5.22 & $^{**}$ & 0.010 & \\
$+220$ &  9.44 & $^{***}$& 0.000453 &
        & 0.50 & (n.s.) & 0.613 &
        & 0.49 & (n.s.) & 0.617 & \\
\hline
Sig./8 & \multicolumn{2}{c}{7/8} & &
        & \multicolumn{2}{c}{0/8} & &
        & \multicolumn{2}{c}{4/8} & \\
\end{tabular}
\end{ruledtabular}
\end{table*}

The terminal Admin--Geo divergence values used in the discussion are summarized in the following table.

\begin{table}[H]
\footnotesize
\setlength{\tabcolsep}{4pt}
\caption{\label{tab:divergence}%
  Deterministic Admin--Geo divergence
  $\DHH(d) = \Hgeo(d) - \Hadm(d)$
  at terminal decades.
  $\DHH = 0$ for all $d \leq 180$~CE
  (networks are identical by construction).
  Values correspond to the current deterministic run.}
\begin{ruledtabular}
\begin{tabular}{r D{.}{.}{4} D{.}{.}{4} D{.}{.}{4}}
$d$ (CE)
  & \multicolumn{1}{c}{$\Hgeo$}
  & \multicolumn{1}{c}{$\Hadm$}
  & \multicolumn{1}{c}{$\DHH$} \\
\hline
$180$ & 2.2994 & 2.2994 & 0.0000 \\
$190$ & 3.2720 & 1.9230 & 1.3490 \\
$200$ & 3.1649 & 1.4438 & 1.7211 \\
$210$ & 3.0709 & 1.4464 & 1.6245 \\
$220$ & 3.0650 & 0.0000 & 3.0650 \\
\end{tabular}
\end{ruledtabular}
\end{table}

The Guo-level connected-component check is included here because it clarifies why the Three Kingdoms signature appears in $\beta_1$ rather than $\beta_0$.

\begin{table}[H]
\footnotesize
\setlength{\tabcolsep}{4pt}
\caption{\label{tab:b0_trajectory}%
  $\beta_0$ (connected components) for
  $\Hgeo$ (Guo level) in the terminal phase,
  confirming $\beta_0 = 1$ throughout.
  Despite Channel~5 inter-domain multipliers
  of up to $25\times$, the Guo-level LCC
  remains globally connected.}
\begin{ruledtabular}
\begin{tabular}{r r r D{.}{.}{4}}
$d$ (CE) & $|\Vgeo(d)|$ & $\beta_0^{\,\mathrm{Geo}}$
          & \multicolumn{1}{c}{$\Hgeo$} \\
\hline
$180$ &  70 & 1 & 2.2994 \\
$190$ & 386 & 1 & 3.2720 \\
$200$ & 386 & 1 & 3.1649 \\
$210$ & 387 & 1 & 3.0709 \\
$220$ & 392 & 1 & 3.0650 \\
\end{tabular}
\end{ruledtabular}
\end{table}

The county-level $\beta_0$ table follows the Guo-level check to show the complementary atomization signal at finer spatial resolution.

\begin{table}[H]
\footnotesize
\setlength{\tabcolsep}{4pt}
\caption{\label{tab:b0_xian}%
  $\beta_0$ and $H$ for the county-level
  geographic network ($H_\mathrm{Geo,Xian}$,
  2,337~Xian nodes, KNN $k=4$) in the
  terminal phase.
  The baseline fragmentation
  ($\beta_0^\mathrm{Xian} \approx 294$ at $d=180$~CE)
  reflects the KNN topology structure;
  the jump to $\beta_0 = 1{,}174$ at $d=190$~CE
  is the historical signal of sub-imperial atomization.
  $N_\mathrm{LCC}$: nodes in the largest
  connected component (on which $H$ is computed).
  Note that $\beta_0^\mathrm{Guo} = 1$ throughout
  the same period (Table~\ref{tab:b0_trajectory}).}
\begin{ruledtabular}
\begin{tabular}{r r r r D{.}{.}{4}}
$d$ (CE) & $|\Vgeo^\mathrm{Xian}(d)|$
  & $\beta_0^\mathrm{Xian}$
  & $N_\mathrm{LCC}$
  & \multicolumn{1}{c}{$H_\mathrm{Geo,Xian}$} \\
\hline
$180$ &   887 &   294 & 88  & 1.6277 \\
$190$ & 2,197 & 1,174 & 100 & 1.4333 \\
$200$ & 2,239 & 1,151 & 174 & 1.9174 \\
$210$ & 2,266 & 1,134 & 250 & 2.8056 \\
$220$ & 2,337 & 1,199 & 215 & 2.4675 \\
\end{tabular}
\end{ruledtabular}
\end{table}

The updated bootstrap information is placed with the divergence results so the deterministic values and the available v7.0-b01 uncertainty information can be read together.

\begin{table}[H]
\footnotesize
\setlength{\tabcolsep}{4pt}
\caption{\label{tab:bootstrap}%
  Updated deterministic and bootstrap summary for the terminal decades
  using the current v7.0-b01 values available at this stage.
  $H_\mathrm{det}$ denotes the deterministic run; $\bar H_\mathrm{boot}$
  denotes the bootstrap mean. The updated Geo confidence interval at
  $d=190$~CE is shown explicitly because it is used in the text;
  the remaining v7.0-b01 confidence intervals should be inserted when
  the complete bootstrap summary is available.}
\begin{ruledtabular}
\begin{tabular}{r D{.}{.}{4} D{.}{.}{4} D{.}{.}{4} l}
$d$ (CE)
  & \multicolumn{1}{c}{$\Hadm^{\rm det}$}
  & \multicolumn{1}{c}{$\bar H_{\rm Admin}^{\rm boot}$}
  & \multicolumn{1}{c}{$\Hgeo^{\rm det}$}
  & \multicolumn{1}{c}{available Geo CI$_{95}$} \\
\hline
$180$ & 2.2994 & 2.2250 & 2.2994 & --- \\
$190$ & 1.9230 & 1.9580 & 3.2720 & $[3.232,\,3.390]$ \\
$200$ & 1.4438 & 1.5886 & 3.1649 & --- \\
$210$ & 1.4464 & 1.0682 & 3.0709 & --- \\
$220$ & 0.0000 & 0.4890 & 3.0650 & --- \\
\end{tabular}
\end{ruledtabular}
\end{table}

\begin{figure}[H]
    \centering
    \includegraphics[width=0.7\linewidth]{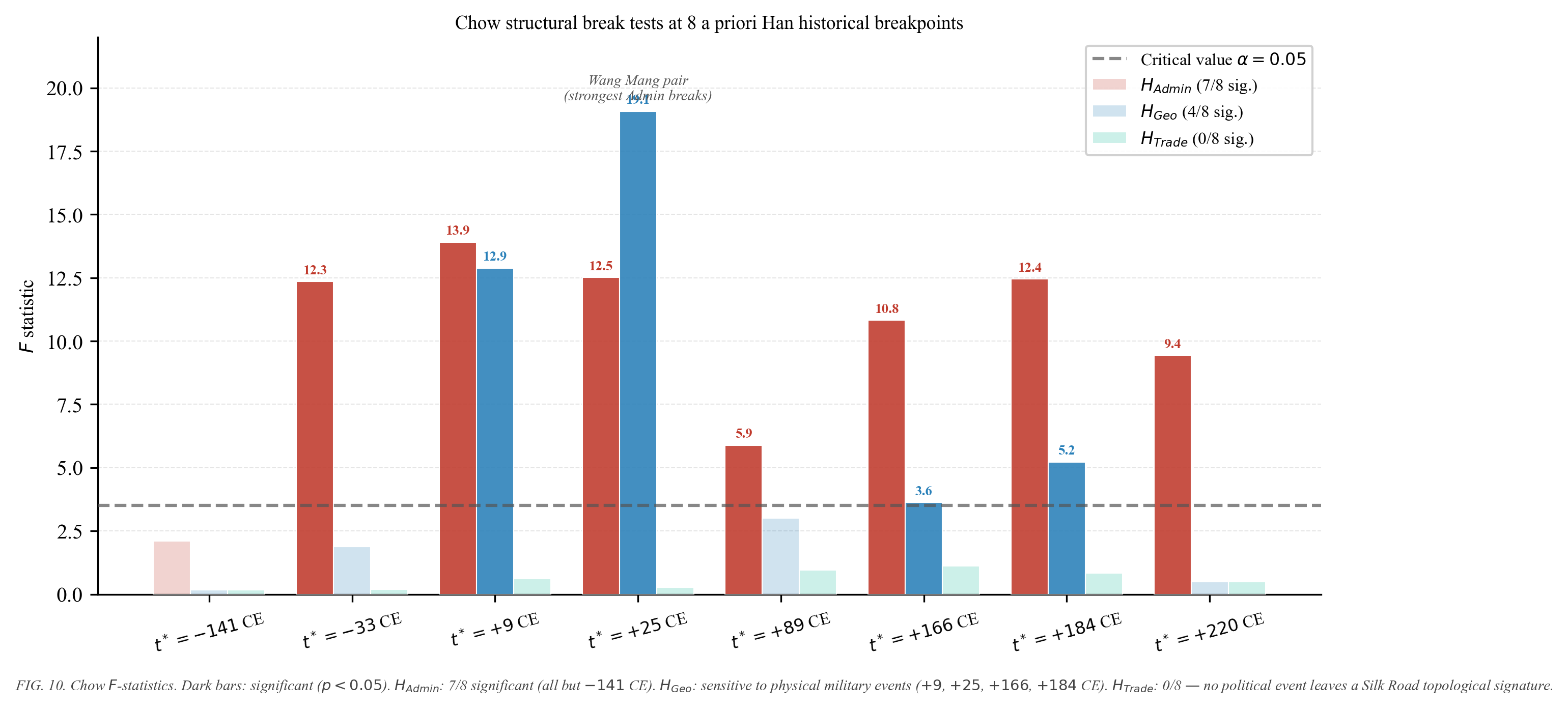}
    \caption{\label{fig:chow_bars}%
    Chow structural break $F$-statistics
    for $\Hadm$ (red), $\Hgeo$ (blue),
    and $\Htrd$ (green) at the eight
    Han historical breakpoints.
    The pattern of significant vs.\ non-significant
    breakpoints discriminates between administrative
    (Admin-only) and physical territorial
    (Geo + Admin) events.}
\end{figure}


The threshold-crossing summary is kept next to the conclusions because it condenses the final status of the three network observables.

\begin{table}[H]
\footnotesize
\setlength{\tabcolsep}{4pt}
\caption{\label{tab:hstar_summary}%
  Summary of $\Hstar = 0.5241$ crossing
  for the three Han network series.
  $\tstar$: decade at which the series first
  crosses $\Hstar$ from above.
  $\gap = 220 - \tstar$: gap between
  crossing and formal dissolution.
  $H_{\rm final}$: value at 220~CE.
  $\Hadm$-only crosses before dissolution;
  $\Hgeo$ never crosses;
  $\Hcomb$ crosses at dissolution by calibration.}
\begin{ruledtabular}
\begin{tabular}{lccccc}
Network
  & $H_{\rm final}$
  & $\tstar$ (CE)
  & $\gap$ (yr)
  & $|\hat\beta_1^G|$ (yr$^{-1}$)
  & $R^2_G$ \\
\hline
$\Hadm$    & 0.000 & ${\sim}216$ & ${\sim}4$ & 0.0261 & 0.689 \\
$\Htrd$    & 1.228 & never       & ---        & 0.0020\ns & 0.21 \\
$\Hgeo$    & 3.065 & never       & ---        & 0.0169 & 0.636 \\
$\Hcomb$   & 0.524 & 220         & 0          & 0.0256 & 0.953 \\
\end{tabular}
\end{ruledtabular}
\end{table}
 
\section{\label{sec:discussion}Discussion and conclusions}


We applied persistent homology to the Han imperial network
($206$~BCE--$220$~CE), constructing three complementary weighted
graphs from geospatial and historical datasets (CHGIS~v6,
a $1~\mathrm{km/pixel}$ DEM, and a georeferenced Silk~Road dataset)
modulated by five historical friction channels.
The principal conclusions are:

\begin{enumerate}

  \item \textbf{Three-network are necessary.}
    $\Hadm$, $\Hgeo$, and $\Htrd$ are topologically irreducible observables giving us information on complementary properties of the Han empire. 

  \item \textbf{Early-warning observables detect the
    collapse 45--50 years before formal dissolution.}
    Three independent indicators---Wasserstein velocity $\dot{W}_2$,
    correlation length $\xi$, and the Integrated Change Tracker
    (ICT)---provide a quantified collapse chronology
     independent of the entropy series.

  \item \textbf{Han administrative collapse recorded with
    high statistical confidence.}
    $\Hadm$ declines at $-0.0261~\text{yr}^{-1}$
    ($R^2 = 0.689$, $p = 0.0108$) in phase~G, reaching
    $\Hadm(220) = 0$ via a condensation funnel from 24~bars
    at $d = 0$~CE to a single continental-scale loop at 220~CE.
    Seven of eight Chow test breakpoints are significant.

  \item \textbf{$\Hgeo$ increases during administrative collapse:
    the internal-fragmentation signature.}
    Identical to $\Hadm$ until 184~CE, $\Hgeo$ diverges positively
    after the Yellow~Turban Rebellion
    (slope $+0.0169~\text{yr}^{-1}$, $R^2 = 0.636$, $p = 0.018$),
    reaching $\Hgeo(220) = 3.065$.
    The cross-network Wasserstein distance
    $W_\times(\mathrm{Admin},\mathrm{Geo})$ jumps from exactly zero
    to $737$ in a single decade at $d = 190$~CE,
    and the correlation length $\xi_\mathrm{Geo}$ diverges
    monotonically to $55\times$ the pre-crisis baseline by 220~CE.
    At the county level (2,337~Xian nodes), $\beta_0$ explodes
    from 294 to 1,174 at $d = 190$~CE, revealing a prior
    atomization phase before the Three~Kingdoms consolidated.

  \item \textbf{The Three~Kingdoms appear as emergent $\beta_1$
    cycles \emph{before} $\Hadm$ reaches zero.}
    Three long-lived $\beta_1$ cycles with a signal-to-noise
    gap of $10.9\times$ over the background emerge at $d = 190$~CE---
    thirty years before the formal Three~Kingdoms declarations
    (220--222~CE).
    The three cycles emerge from geographic node locations,
    LCP-derived costs, and the historically motivated post-184~CE
    fragmentation model, corresponding to the Wei, Shu, and Wu domains.
    This suggests that TDA can recover politically meaningful geographic domains
    when topographic structure and historically motivated friction channels are incorporated.

  \item \textbf{The Silk~Road is a structural invariant
    of the Han collapse.}
    $\Htrd$ shows no significant structural breaks at any of the
    eight historical breakpoints (0/8), including the final
    dissolution.
    Long-distance trade circuits maintain topological stability
    across political disruptions, consistent with the historical
    record of Silk~Road continuity into the Three~Kingdoms period.

\end{enumerate}

The results demonstrate that topological data analysis,
applied to appropriately decomposed historical networks,
can distinguish mechanistically distinct collapse modes,
identify topological precursors of collapse decades before
formal dissolution, and predict the spatial structure of
successor polities without historical knowledge of the
eventual outcome.

\begin{acknowledgements}
The CHGIS dataset is made available by [Bol \& Ge, Harvard CGIS].
TDA computations were performed using the Gudhi library~\cite{gudhi}.We acknowledge financial support from SECIHTI and SNII (M\'exico)
\end{acknowledgements}



\end{document}